\documentclass[manuscript,screen,table]{article}

\usepackage[a4paper, margin=1in]{geometry}
\usepackage{graphicx}
\usepackage{authblk}
\usepackage[hyphens]{url}
\usepackage{multirow}
\usepackage{rotating}
\usepackage{booktabs}
\usepackage[inline]{enumitem}
\PassOptionsToPackage{xcolor}{table}
\usepackage{array}
\newcolumntype{P}[1]{>{\centering\arraybackslash}p{#1}}
\usepackage{calrsfs}
\usepackage{amsmath,amssymb,amsfonts}
\usepackage{algorithmic}
\usepackage{verbatim}
\usepackage{xcolor}
\usepackage{tikz}
\usepackage[noorphans,font=itshape]{quoting}
\usepackage{pgfplots}
\pgfplotsset{compat=1.12}
\definecolor{mobile}{RGB}{255,137,0}
\definecolor{fsdb}{RGB}{21, 96, 122}
\definecolor{blockchain}{RGB}{86, 248, 180}
\definecolor{iot}{RGB}{199,30,29}
\definecolor{cloud}{RGB}{0,144,118}
\definecolor{multimedia}{RGB}{24,161,205}
\definecolor{networks}{RGB}{255,233,114}
\definecolor{misc}{RGB}{197, 216, 227}

\newcommand{\myparagraph}[1]{\vspace{0.1cm}\noindent{\it \textbf{#1}.}}

\providecommand{\keywords}[1]{\textbf{\textit{Index terms---}} #1}
\begin{document}

\title{Research trends, challenges, and emerging topics of digital forensics: A review of reviews}

\author[1,2]{Fran Casino}
\author[3]{Tom Dasaklis}
\author[4]{Georgios Spathoulas}
\author[5]{Marios Anagnostopoulos}
\author[6]{Amrita Ghosal}
\author[7]{Istv\'an Bo\"ro\"cz}
\author[8]{Agusti Solanas}
\author[9]{Mauro Conti}
\author[1,2]{Constantinos Patsakis}

\affil[1]{University of Piraeus, Greece}
\affil[2]{Athena Research Center, Greece}
\affil[3]{Hellenic Open University, Greece}
\affil[4]{University of Thessaly, Greece}
\affil[5]{Aalborg University, Denmark}
\affil[6]{University of Limerick, Ireland}
\affil[7]{Vrije Universiteit Brussels, Belgium}
\affil[8]{Universitat Rovira i Virgili, Spain}
\affil[9]{University of Padua, Italy}
\date{}
\maketitle

\begin{abstract}
Due to its critical role in cybersecurity, digital forensics has received significant attention from researchers and practitioners alike. The ever increasing sophistication of modern cyberattacks is directly related to the complexity of evidence acquisition, which often requires the use of several technologies. To date, researchers have presented many surveys and reviews on the field. However, such articles focused on the advances of each particular domain of digital forensics individually. Therefore, while each of these surveys facilitates researchers and practitioners to keep up with the latest advances in a particular domain of digital forensics, the global perspective is missing. Aiming to fill this gap, we performed a qualitative review of reviews in the field of digital forensics, determined the main topics on digital forensics topics and identified their main challenges. Our analysis provides enough evidence to prove that the digital forensics community could benefit from closer collaborations and cross-topic research, since it is apparent that researchers and practitioners are trying to find solutions to the same problems in parallel, sometimes without noticing it.
\end{abstract}



\section{Introduction}

According to Edmond Locard's exchange principle, in every crime, the perpetrator will alter the crime scene by bringing something and leaving something else. Therefore, these changes can be used as forensic evidence. While this principle is relatively straightforward, it is difficult in many cases to apply. This is why Locard introduced forensics labs in Law Enforcement Agencies (LEAs) over the first decade of the 20th century.

While procedures that resemble digital forensics are mentioned in computer science literature quite early, the domain was not fully defined until 1980s when it started to gain attention. The introduction of the IBM PC generalised the use of computing machines; thus, more interest was focused on digital evidence and many people came together and created a digital forensics community, which eventually became more formal in 1993 when the FBI hosted the First International Conference on
Computer Evidence \cite{pollitt2010history}. Initially, the main activity was examining standalone computers to recover deleted or destroyed files from the disks. However, since the early 2000s, the digital forensics domain has expanded steadily, maturing along with regulations. Nowadays, users tend to utilise multiple digital devices and access tenths of digital services per day. The digital footprint of our everyday life has become enormous, and accordingly the probability that illegal activities leave digital evidence behind is very high. The need for forensic investigators has increased, and this have led to multiple academic education and certification programs related to digital forensics \cite{5428493}. Additionally, the complexity of the tasks to be carried out and the required compliance with law and courts' regulations has led to the establishment of strict protocols and procedures to be followed \cite{marshall2010quality,chen2005standardizing}. The continuous appearance of new forms of cybercrime also requires adaptive investigation process models, new technology, and advanced techniques to deal with such incidents.

Beyond the rise of cybercrime, where the evidence is expected to be digital, digital evidence is underpinning almost all modern crime scenes. For instance, mobile devices have become a primary source of digital evidence as almost all our communications are performed through them. In fact, according to EU\footnote{\url{https://ec.europa.eu/commission/presscorner/detail/en/MEMO_18_3345}}, the bulk of criminal investigations (85\%) involve electronic evidence.
Thus, emails, cloud service providers, online payments, and wearable devices are often used to extract digital evidence in various circumstances.

\noindent{}\textbf{Motivation and contribution:}
\label{sec:motivation}
Digital evidence has become a norm and underpins most modern crime investigations. However, there are digital evidence to which different methods and methodologies apply. Some principles may remain the same; however, they cannot be applied to all types of evidence. For instance, collecting evidence from the Cloud bears no resemblance to IoT forensics or image forensics. This has led to a huge amount of research, which addresses the challenges raised in each domain individually, with the bulk of the work devoted to the development of novel tools and algorithms to extract digital evidence and intelligence from heterogeneous sources. Currently, investigators devote many efforts to provide a systematic overview of the literature and the advances in each domain, with focused surveys and reviews. Despite the importance of these surveys, the overall picture is still missing. Each of these surveys is focused on a specific domain and, as a result, common issues, challenges and methods are not identified. Moreover, research directions and approaches, that could be applied in several domains, remain explored in a topic-wise manner, lacking interoperability, and denoting a lack of collaboration between researchers in different forensics domains. 

We sustain that the above is a serious gap in current literature, and we aim to fill it in this article. To this end, we present a review of reviews in the field of digital forensics. We collect all relevant surveys and reviews in the field of digital forensics, analyse them, and answer a set of research questions, listed in Table \ref{tab:attacks-priv-table}, by performing the following actions:
\begin{itemize}
    \item Analysing the current state of the art and practice, and identifying the challenges of each domain individually.
    \item Assessing whether the current state of the art is aligned with the technological evolution in digital forensics.
    \item Using the previously collected information to identify common issues, gaps, best strategies and key focus areas in digital forensics, trying to span across different domains.
    \item Assessing technological advances to highlight emerging challenges in digital forensics.
\end{itemize}
Our work goes beyond a thorough taxonomy. We analyse several dimensions of digital forensics, covering frameworks and process models, standardisation, readability and reporting, as well as legal and ethical aspects.
To the best of our knowledge, this is the first review of reviews covering the state of the art in digital forensics and showcasing the actual state of practice from a global perspective.




\begin{table*}[h]
\renewcommand{\arraystretch}{1.1}
 \centering
 \footnotesize
 \rowcolors{2}{gray!25}{white}
 \begin{tabular}{p{4.5cm}p{9.75cm}p{2.25cm}}
\toprule \textbf{Research Question} & \multicolumn{1}{c}{\textbf{Objective}} & \textbf{Relevant Sections}\\
 \midrule
 What is the current state of practice and research trends? & The objective is to discover in what topics researchers and practitioners are devoting more efforts, so that we can identify both the research trends and the topics that require more support. Furthermore, it will streamline common solutions and practices that can be fostered by other domains. & 2, 3 and 4 \\
 Which are the current challenges in digital forensics? & Since digital forensics is applied in multiple contexts and to different technologies, our aim is to extract the challenges in both local and global perspective to provide a comprehensive overview. This will highlight the particularities of each domain but also stress their commonalities. &  3 and 4 \\
 Is the current state of the art aligned with technological evolution in digital forensics? & The intention is to analyse whether the actual state of the art, in terms of e.g., technologies, legislation and standards, is sufficient to cope with modern cybercrime. This can serve as a road map for tool development, prioritisation standardisation actions etc. &  3, 4 and 5 \\
 What strategies should be used to deal with identified challenges? & According to the knowledge extracted from the state of the art, our plan is to identify the pain points of the actual state of practice and leverage a gap analysis to provide fruitful strategies. &  4 and 5 \\
 Based on technological advances and trends, which are the challenges that digital forensics will face in the upcoming future? & In this question, our goal is to identify characteristics and critical issues in emerging technologies that may hinder digital investigations in the near future. Timely identifying these issues and prioritising R\&D actions will significantly decrease their potential impact.  &   5 \\
 \bottomrule
 \end{tabular}
 \caption{Summary of research questions and the corresponding sections devoted to answer them.}
 \scriptsize
 \label{tab:attacks-priv-table}%
\end{table*}%

The remainder of the article is organized as follows: Section \ref{sec:methodology} details our research methodology, providing a descriptive analysis of the retrieved literature, which is then complemented with a taxonomy of digital forensics in Section \ref{sec:taxonomy}. Section \ref{sec:digitalforensics_mps} analyses the current state of practice regarding forensic methodologies and their phases, standards, and ethics. Relevant open issues, trends, and further research lines are discussed in Section \ref{sec:discussion}. The article concludes in Section \ref{sec:conclusions} with some final remarks.


\section{Research Methodology}
\label{sec:methodology}

In recent years, academic publishing has significantly increased both in terms of volume and speed. At the same time, new channels for publication, such as conference proceedings, open archives and numerous scientific journals, are rapidly expanding, thus allowing today's researchers to publish their work in a multitude of venues \cite{fire_over-optimization_2019}. According to recent studies, approximately 22 new systematic reviews are published daily \cite{hunt_introduction_2018}. New methodological approaches for synthesising this evidence have been developed to keep up with the proliferation of systematic reviews across disciplines. Besides, conducting reviews of existing systematic reviews has become a logical next step in providing evidence in domains where a growing number of systematic reviews is available. Overviews or umbrella reviews are most commonly used to bring together, appraise, and synthesise the results of related systematic reviews when multiple systematic reviews on similar or related topics already exist \cite{hunt_introduction_2018,mckenzie_overviews_2017}. Therefore, a review of reviews or an umbrella review compiles evidence from multiple reviews or survey papers into a single document. Syntheses of previous systematic reviews are known by a variety of names, one of which is an umbrella review. Other descriptions include the terms (``review of reviews,'' ``systematic review of reviews,'' ``review of systematic reviews,'' ``overviews of reviews,'' ``summary of systematic reviews,'' ``summary of reviews,'' and ``synthesis of reviews'') \cite{Aromataris2015132}.

\begin{figure}[th!]
\centering
\includegraphics[width=.7\columnwidth]{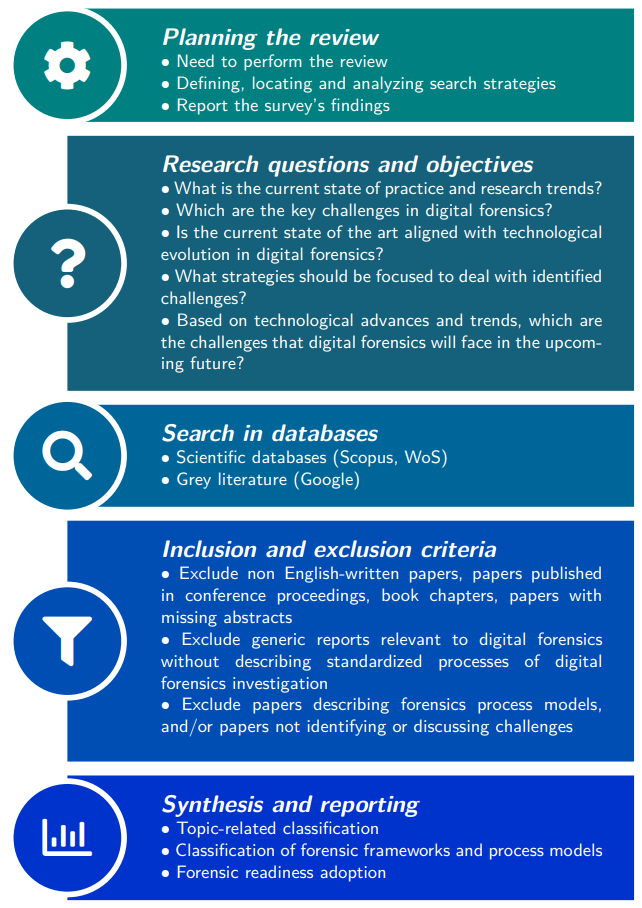}
\caption{Detail of the research methodology steps.} \label{fig:classification_research}
\end{figure}

Despite their growing popularity, no standardized reporting guidelines currently exist for umbrella reviews. However, various multidisciplinary teams around the globe work together to develop relevant standardized reporting guidelines that will soon be available \cite{pollock_preferred_2019}. In our case, we rely upon an entirely systematic way to conduct our umbrella review. In particular, we have used various features of the approach presented in \cite{denyer2009producing} to conduct our review of reviews and provide a transparent, reproducible and sound overview of the scientific literature on digital forensics from a global perspective. Our review protocol consists of five steps, as shown in Figure \ref{fig:classification_research}: 1) Planning the review 2) Defining research questions 3) Searching literature databases 4) Applying inclusion and exclusion criteria and 5) Synthesising and reporting the results of the literature analysis.

\subsection{Search strategy}

As previously stated, our overall survey process is based on several predefined research questions relevant to the digital forensics literature. We conducted extensive research addressing the various technical/functional/security challenges of the digital forensics literature guided by these research questions. To this end, we performed a systematic literature search without time constraints in May 2021 which was subsequently updated in November 2021. The main search engines used were Web of Science (WoS), Scopus and Google. Scopus and WoS were used to locate all scientific-related literature due to their multidisciplinary coverage and scope \cite{pranckute2021web}, while Google was used to locate relevant standards and best practices (grey literature). We queried Scopus and WoS using the terms "\textit{digital forensics and review or survey}" in the title, keywords, and abstract of all articles. It is worth noting that first bulk search query yielded 536 unique results (combining both sources). 

Electronic searches using Google also turned up relevant \emph{grey literature}, such as unpublished research commissioned by governments or private/public institutions. In particular, we looked at the first 200 Google results for the queries \textit{digital forensics and reviews} and \textit{digital forensics and surveys} to find the published grey literature. 
It is worth noting that we used Google searches as a supplement to our primary search strategy (especially for streamlining the assessment), and Scopus and WoS were our primary source for finding scientific-related literature. Furthermore, compared to the bibliography retrieved from Scopus and WoS, the total number of documents retrieved from Google was relatively low.

We discovered additional studies using the so-called snowball effect (backward and forward), which involved searching the references of key articles and reports for additional citations \cite{vom2015standing}. For instance, additional grey literature was discovered by manually searching the reference lists in several reports, particularly research and committee reports or policy briefs from private and public sector institutions/organizations. For this study, we take into consideration 109 research papers and 51 reports. The 109 papers are used for identifying relevant challenges/trends across different digital forensics domains (see Section \ref{sec:taxonomy} ). The 51 reports were used to derive further insights about the state of practice regarding digital forensics methodologies, practices and standards, as well as discussing future trends and open challenges from a policy perspective (see sections \ref{sec:digitalforensics_mps} and \ref{sec:discussion}). 

\subsection{Selection of studies}

We used various pre-defined exclusion and inclusion criteria as described in Table \ref{tab:selection-criteria} to assess the eligibility of the retrieved literature; both academic and grey. Some exclusion criteria were used before introducing the literature into the bibliographic manager (language, subject area and document type restrictions). It is also worth noting that we have only examined review papers and reports written in English.

Our overall selection process steps are the following: (i) We initially evaluated the relevance of the titles of all scientific articles and reports. Articles/reports fulfilling one of the exclusion criteria were removed from the analysis and sorted according to the reason for their removal, (ii) In the sequence, we evaluated the relevance of all paper abstracts and report introduction sections (grey literature). Articles and/or reports that met one of the defined exclusion criteria were excluded from the analysis, and we documented the reason for exclusion, (iii) We also did a full-text reading, and some additional articles/reports were excluded and sorted by reason of exclusion during this step. We resolved any potential disagreements among authors about the relevance of the retrieved articles/reports through discussion until reaching a unanimous consensus. We omitted several studies because they were not reviews or surveys (for example, papers relevant to financial forensics investigation, business forensics). We also discarded from the analysis articles that did not meet the inclusion criteria.

\begin{table*}[t]
\centering
\renewcommand{\arraystretch}{1.1}
\scriptsize
\resizebox{\linewidth}{!}{
\begin{tabular}{l|p{.25\textwidth}|p{.25\textwidth}|p{.25\textwidth}}
\hline
\multicolumn{1}{c|}{\textbf{Selection  criteria}} & \multicolumn{2}{c|}{\textbf{Scientific  database}} & \multicolumn{1}{c}{\textbf{Grey literature}} \\ \hline
\multirow{3}{*}{\textbf{Inclusion}} & \multicolumn{2}{p{3in}|}{Only peer-reviewed scientific research papers (including articles in press, written in English)} & Industry reports, committee reports, policy briefs (written in English) \\ \cline{2-4}
 & \multicolumn{2}{l|}{Without time-frame restrictions} & Without time-frame restrictions \\ \hline
\multirow{6}{*}{\textbf{Exclusion}} & Before import to the bibliographic manager & Non English-written papers, papers published in conference proceedings, book chapters, papers with missing abstracts etc. & \multirow{6}{1.5in}{Generic reports relevant to digital forensics without describing standardized processes of digital forensics investigation.} \\ \cline{2-3}
 & During  abstract screening & Papers belonging to other discipline than digital forensics&   \\ \cline{2-3}
 & During full-text reading & Papers describing forensics process models, and/or papers not identifying or discussing challenges &   \\ 
\hline
\end{tabular}
}
\caption{Selection criteria of the retrieved
literature.}

\label{tab:selection-criteria}
\end{table*}

\subsection{Analysis and reporting}

All articles and/or reports that met the inclusion criteria were analyzed (in emerging themes) using a qualitative analysis software (MAXQDA11). The authors carried out the thematic content analysis independently. We applied various qualitative analysis methods (such as narrative synthesis and thematic analysis) to classify and synthesise the extracted data in a sound and comprehensive manner. The results of our analysis are presented in sections \ref{sec:taxonomy} and \ref{sec:digitalforensics_mps}.

\subsection{Bibliographic analysis}

\begin{figure*}
\centering
\resizebox{\linewidth}{!}{
\includegraphics[width=\textwidth]{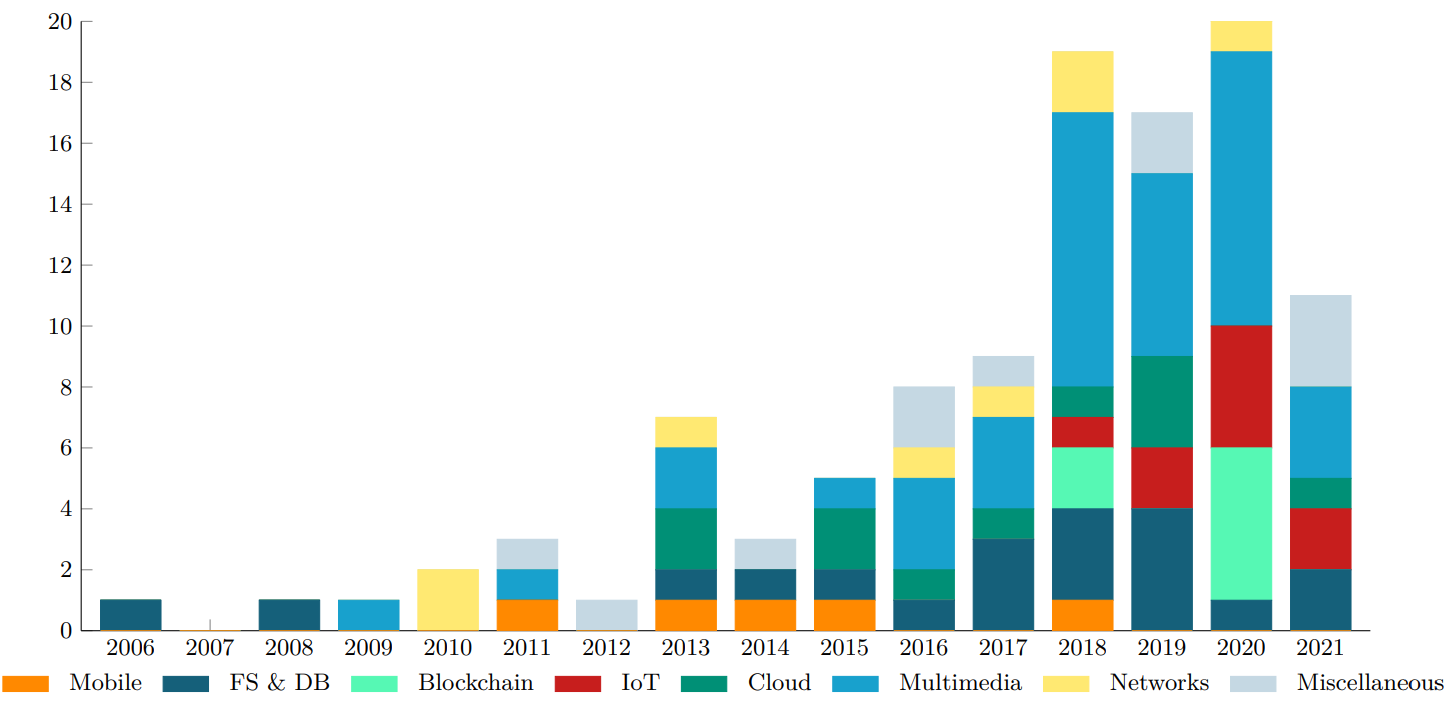}
}
\caption{Year-wise analysis of the selected literature per domain.}
\label{fig:classification_bib}
\end{figure*}

In this section, we present a descriptive analysis of the scientific papers included in the challenges-based and domain-specific classification (see Figure \ref{fig:classification_bib}). The descriptive analysis includes 109 research papers published from 2006 until the end of November 2021. The purpose of the descriptive analysis presented is three-fold: \begin{enumerate}
\item It enhances the statistical description, aggregation, and presentation of the constructs of interest or their associations of the relevant literature (publications per year and domain etc.).
\item It contains insights to current research trends in the area of digital forensics and a critical discussion of the challenges identified. It, therefore, supports the classification structure presented in Section \ref{sec:taxonomy}
\item It allows us to visually demonstrate the diverse research approaches used up to this point in the scientific literature regarding the proliferation of digital forensics review papers.
\end{enumerate}

The distribution of publications over time is depicted in Figure \ref{fig:classification_bib}. In particular, Figure \ref{fig:classification_bib} shows a year-by-year analysis of the selected papers. It is worth noting that the number of publications has increased significantly after 2017. Until the end of 2017, there were only about 38 review papers addressing issues of digital forensics. However, from 2017 onwards, the number of reviews published in the scientific literature has risen to nearly 70. As a result, over the last four years, research in the area of digital forensics has slowly but steadily increased. This upward trend reflects the key public and policy impact of digital forensics nowadays.

Figure \ref{fig:classification_bib} also shows the domain-specific distribution of the 109 review papers included in our analysis. It is worth noting that we have identified seven (7) prevalent areas of research interest in digital forensics: Blockchain, Cloud, Filesystem and databases, Multimedia, IoT, Mobile, Networks. Multimedia forensics attracts most of the current digital forensics research (38 out of the 109 review papers), followed by Filesystem and database forensics papers (18 out of 109). Both streams justify that the widespread use of mobile devices with lower-cost storage and increased bandwidth has resulted in a massive generation of multimedia-related content. Furthermore, various miscellaneous review papers (applications that do not fit into any of the above categories) demonstrate the digital forensics multidisciplinary nature. These multidisciplinary review papers represent research conducted in areas such as social media, smart grid, unmanned aerial vehicles and etc.

\section{Taxonomy of Challenges-based Digital Forensics Research}
\label{sec:taxonomy}

\begin{figure*}[th!]
\centering
\includegraphics[width=.75\textwidth]{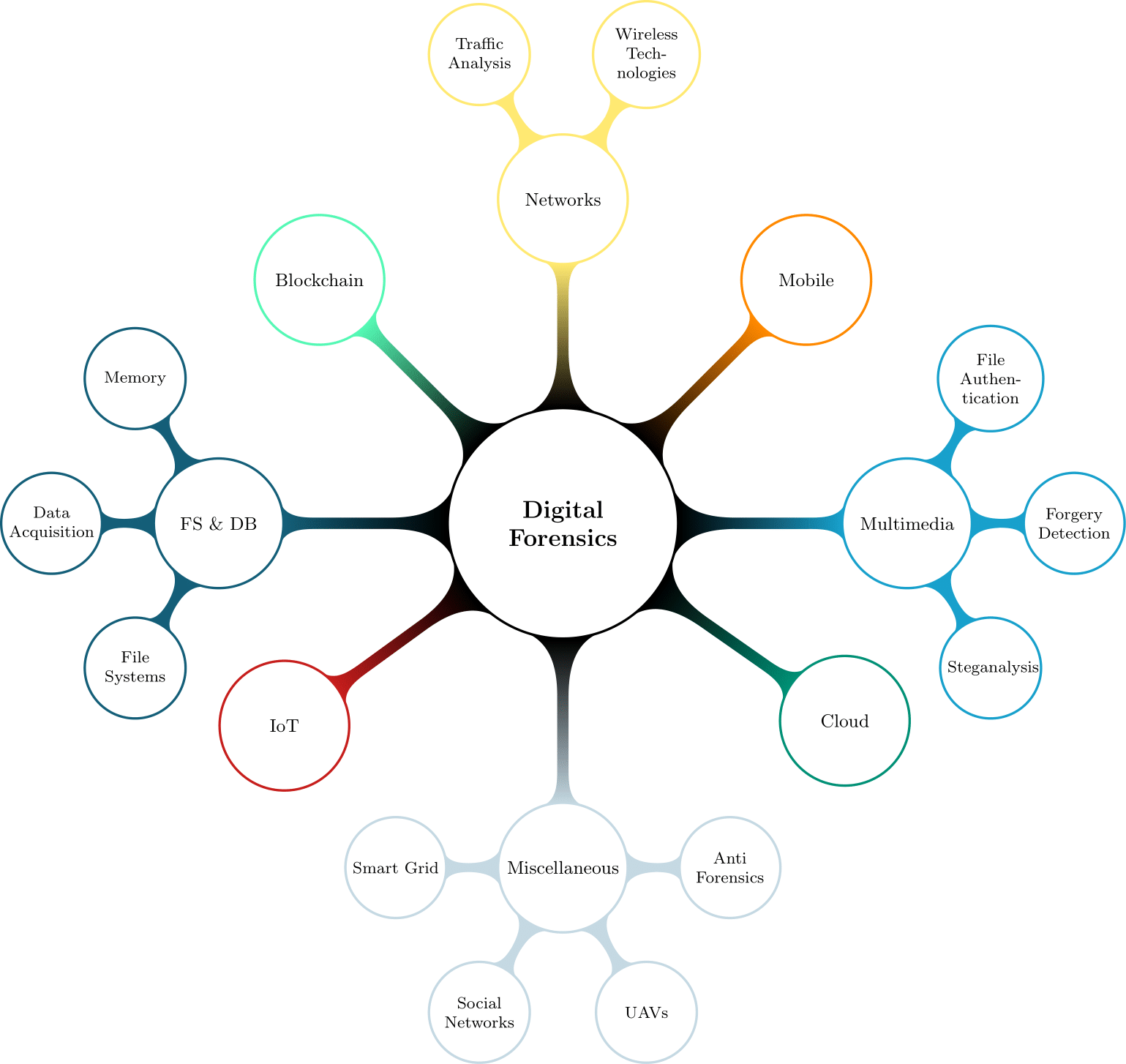}
\caption{Challenges-based and domain-specific mindmap abstraction of digital forensics topics identified in the literature.} \label{fig:classification}
\end{figure*}

In this section, we summarise the surveys/literature reviews collected following a rigorous statistical methodology based on the literature, as described in Section \ref{sec:methodology}. The topics of this classification have been systematically selected according to the contents of reviewed literature, and thus reflect the digital forensics research landscape and illustrates with high fidelity the heterogeneity of digital forensic solutions. The classification of digital forensics topics is graphically represented in Figure \ref{fig:classification}. In each case, we discuss the main limitations and challenges proposed in the literature. More precisely, we extract the challenges at a research field domain level (i.e., we group in a higher hierarchical level, when possible, the limitations of the methods presented in the surveys) to give a more comprehensive perspective and to enable further cross-topic comparisons in Section \ref{sec:aggregated}.

\subsection{Cloud}
\label{sec:cloud_taxonomy}

Researchers, as well as government agencies, have thoroughly explored many of the challenges in cloud forensics, though some challenges still remain to be addressed. For example, the diversity of embedded OSs with shorter product life cycles, as well as the numerous smartphone manufacturers around the world present, are challenges in this research area. In the literature, we can find research works that have addressed challenges in cloud forensics and their solutions from different perspectives. Purnaye et al. \cite{purnaye2021comprehensive} explored the different dimensions of cloud fornesics and categorised the main challenges of this topic. Alex et al.~ \cite{alex17} discussed challenges in cloud forensics related to data acquisition, logging, dependence on cloud service providers, chain of custody, crime scene reconstruction, cross border law and law presentation. Khanafseh et al.~ \cite{khanafseh19} pointed out several challenges in cloud forensics, such as the unification of logs format, missing terms and conditions in Service Level Agreement (SLA) regarding investigations where service level agreement is the main point and condition between the user and the cloud service provider, lack of forensics expertise, decreased access to forensic data and control over forensics data at all level from the customer side, lack of international collaboration and legislative mechanism in cross-nation data access and exchange, and lack of international collaboration and legislative mechanism in cross-nation data access and exchange. Pichan et al.~ \cite{pichan15} considered the Digital Investigative Process (DIP) model \cite{palmer2001road} for describing the challenges emerging at each phase of the digital investigation process and provided solutions for the respective identified challenges. The challenges identified by the authors in cloud forensics are unknown physical location, decentralized data, data duplication, jurisdiction, encryption, preservation, dependence on CSP, chain of custody, evidence segregation, distributed storage, data volatility and integrity. Similar to the works of Khanafseh et al. and Pichan et al., the authors in ~ \cite{ruan13} also identified the challenges in cloud forensics and analyzed them on the basis of their significance. Park et al. \cite{park18} discussed the different challenges within cloud forensic investigations highlighting the relevance of proactive models, and discussing the integration of smart environments to enhance the robustness of forensic investigations. The authors in ~ \cite{simou16} provided a categorization of the cloud forensic challenges based on the cloud forensic process stages. Amminezhad et al.~ \cite{aminnezhad13} described the different challenges in cloud forensics that were addressed by other authors by performing an exploratory analysis. Rahman et al.~ \cite{ab15} broadly classified the existing challenges in cloud forensics, classifying the literature into three categories, namely, multi-tenancy, multi-location and scope of user control. Finally, the authors in~ \cite{manral19} identified and discussed the major challenges that occur at each stage of the cloud forensic investigation, according to well-known forensic flows.

As evident from the large number of publications in literature reviews/surveys, cloud forensics is quite an explored research topic. Despite the considerable amount of research in cloud forensics, there still exist a number of challenges/limitations that need much attention, as discussed by NIST \cite{nist2020cloud}. In Table \ref{tab:cloudforensics}, we present a summary of the extracted challenges in the cloud forensic review/survey articles. From this summary, we observe that there is a dearth of research work focusing on cloud forensic standard tools and technologies in the cloud environment. Also, very limited works have concentrated on pointing out the feasible solutions related to the challenges present in cloud forensics.

\begin{table}[hbt!]
 \begin{footnotesize}
 \rowcolors{2}{gray!25}{white}
 \begin{tabular}{p{.65\linewidth}>{\raggedright\arraybackslash}p{.25\linewidth}}
 \toprule
 {\textbf{Challenge/Limitation}} & \textbf{References}\\
 \midrule
 Update forensic tools to fight novel cybercrime & \cite{manral19,khanafseh19,pichan15,simou16,purnaye2021comprehensive,park18} \\
 Lack of forensic readiness mechanisms and their management & \cite{manral19,alenezi19,alex17,pichan15,park18} \\
 Data management and its fragmentation hinders investigations & \cite{ab15,alenezi19,alex17,khanafseh19,pichan15,ruan13,simou16,purnaye2021comprehensive,aminnezhad13} \\
 Lack of trust and robust chain of custody preservation & \cite{alenezi19,alex17,khanafseh19,pichan15,ruan13,purnaye2021comprehensive,park18} \\
 Lack of jurisdictional mechanisms for confidential data & \cite{alenezi19,ruan13,simou16} \\
 Cross border investigations due to different jurisdictions and laws & \cite{alex17,khanafseh19,pichan15,ruan13,purnaye2021comprehensive} \\
 Lack of training and interoperability between investigators and court & \cite{khanafseh19} \\
 Anti-forensics & \cite{purnaye2021comprehensive}\\
 \bottomrule
 \end{tabular}
 \end{footnotesize}
 \caption{High level extraction of limitations in cloud forensics.}
 \label{tab:cloudforensics}%
\end{table}%

\subsection{Networks}

Data monitoring and acquisition from network traffic are mandatory to prevent most of nowadays cyber-attacks \cite{Pilli2010,ntsc2020,patsakis2020unravelling}, including, but not limited to, Distributed Denial of Service (DDoS), phishing, DNS tunnelling, Man-in-the-middle (MitM) attacks, SQL injection and others \cite{HOQUE2014307,khan2016network}. Regardless of the orchestration mechanism behind them (i.e., single attackers or orchestrated botnets), the analysis and mitigation mechanisms rely on the proper monitoring and analysis of computer network traffic to collect information, evidence and proof of any intrusion detection or vulnerability. For this purpose, several well-known tools exist, such as network forensic analysis tools which provide functionalities such as traffic sniffing, Intrusion Detection Systems (IDS), protocol analysis, and Security Event Management (SEM) \cite{8410366,sharevski2018towards,takahashi2010ieee,khan2016network}. Nevertheless, one of the challenges of network forensics is to achieve accurate and efficient packet analysis in encrypted network traffic since it is far more challenging than the analysis of unencrypted traffic. As authors stated in \cite{SIKOS2020200892,khan2016network}, utilizing machine learning in packet analysis is evolving into a complex research field that aims to address the analysis of unknown features and encrypted network data streams.

Regarding the research and forensics-related surveys tackling such issues, several reviews recall the primary methodologies and tools for network forensic analysis, such as the works seen in \cite{adeyemi2013review,Pilli2010}, yet they were conducted almost a decade ago. Therefore, taxonomies classifying forensic frameworks suitable for Network Forensics are crucial \cite{khan2016network}. An interesting review focusing on the attackers perspective, in terms of attack behaviour and plan identification, as well as prevention mechanisms, can be found in \cite{ahmed2017attack}. Finally, some protocol-oriented reviews, analyzing IEEE 802.11 protocol \cite{takahashi2010ieee}, and more recently, 5G networks \cite{sharevski2018towards}, discuss specific vulnerabilities in their corresponding contexts. In general, the main challenges of network forensics, as identified by the authors in the aforementioned works, are classified in Table \ref{tab:network}.

\begin{table}[t]
 \centering
 \begin{footnotesize}
 \rowcolors{2}{gray!25}{white}
  \begin{tabular}{p{.65\linewidth}>{\raggedright\arraybackslash}p{.25\linewidth}}
\toprule {\textbf{Challenge/Limitation}} & \textbf{References}\\
 \midrule
 Reduce the amount of data required for attack identification & \cite{Pilli2010,SIKOS2020200892,khan2016network} \\
 Heterogeneous data acquisition, integrity and interpretation & \cite{adeyemi2013review,Pilli2010,8410366,sharevski2018towards,khan2016network} \\
 Ubiquitous environments and cross border data & \cite{adeyemi2013review,sharevski2018towards,khan2016network} \\
 Reliable detection of attacks & \cite{adeyemi2013review,Pilli2010,ahmed2017attack,sharevski2018towards,khan2016network} \\
 Increased possibilities of monitoring mechanisms & \cite{takahashi2010ieee,sharevski2018towards, adeyemi2013review} \\
 Efficient and accurate analysis of encrypted traffic & \cite{SIKOS2020200892,khan2016network}\\
 \bottomrule
 \end{tabular}
 \end{footnotesize}
  \caption{High level extraction of challenges in network forensics.}
 \label{tab:network}%
\end{table}%

\subsection{Mobile}
\label{sec:Mobile}
Smartphones and mobile devices may contain valuable information for a plethora of investigation purposes. Mobile forensics (MF) is a sub-branch within the digital forensics domain relevant to the extraction of digital evidence from portable and/or mobile devices. Mobile forensics processes could be broken down into the following three categories: seizure, acquisition, and examination/analysis.

The diversity of embedded OSs with shorter product life cycles, as well as the numerous smartphone manufacturers around the world, stand out as significant challenges in the MF domain \cite{Chu2011}. In general, MF presents various challenges due to a multitude of reasons. For example, in \cite{Barmpatsalou2018} the authors identify the following limitations for successfully carrying out MF investigations: 1) data-related issues (anonymity-enforced browsing and other anonymity services, and the considerable volume of data acquired during an investigation) 2) forensic tools-related issues (MF research approaches have long focused on acquisition techniques, while minor importance was given to the other phases of MF investigative process) 3) device and operating systems diversity 4) security aspects (development of new and more sophisticated anti-forensic methods from the manufacturers) 5) cloud-related issues (current MF tools do not consider cloud aspects, cloud investigation barriers such as access to forensics data due to multi-jurisdictional legal frameworks, forensics data security) and 6) process automation. It is worth noting that MF faces significant challenges concerning the focus of the overall MF processes. For example, it is not clear whether investigation procedures should be model-specific for each device or should be generic enough to form a standardized set of guidelines applicable to forensics procedures \cite{Farjamfar2014}. Another challenge is the need to perform live forensics (mobile device should be powered on) \cite{Barmpatsalou2013}. In addition, an important barrier for actually conducting MF investigations relates to the various networking capabilities of smartphones, which render the overall MF processes difficult to manage, particularly due to the complex structure of the cloud computing environment \cite{Wan2015}. Finally, due to the security measures inherent to modern mobile devices, an investigator must actually break into the device using an exploit that will most likely alter the device data. Clearly, the latter violates the Association of Chief Police Officers (ACPO) principle and introduces numerous procedural issues for a forensic investigation. In Table \ref{tab:mobileforensics}, we provide a classification of MF approaches' current challenges.

\begin{table}[hbt!]
 \centering

 \begin{footnotesize}
 \rowcolors{2}{gray!25}{white}
 \begin{tabular}{p{.65\linewidth}>{\raggedright\arraybackslash}p{.25\linewidth}}
\toprule {\textbf{Challenge/Limitation}} & \textbf{References}\\
 \midrule
Reduced training and data acquisition overheads & \cite{Barmpatsalou2018,Wan2015} \\
Diversity of embedded OSs with shorter product life cycles, multitude of smartphone manufacturers & \cite{Chu2011} \\
Heterogeneous data acquisition and interpretation & \cite{Barmpatsalou2018,Barmpatsalou2013} \\
Update forensic tools to fight novel cybercrime & \cite{Barmpatsalou2018,Wan2015} \\
Strong security mechanisms of mobile devices and anti-forensics& \cite{Barmpatsalou2013,Barmpatsalou2018} \\
The very nature of mobile phones necessitates the adoption of live forensics approaches & \cite{Barmpatsalou2013} \\
Lack of trust and robust chain of custody preservation & \cite{Barmpatsalou2018,Barmpatsalou2013} \\
Lack of device-based standards and procedural guidelines & \cite{Farjamfar2014} \\
Lack of jurisdictional and legal requirements for different investigation scenarios & \cite{Wan2015} \\
 \bottomrule
 \end{tabular}
 \end{footnotesize}
  \caption{High level extraction of limitations in mobile forensics.}
 \label{tab:mobileforensics}
\end{table}%

\subsection{IoT}
\label{sec:IoT}
Although significant in terms of improved data availability and operational excellence, the broad adoption of IoT devices and IoT-related applications have brought forward new security and forensics challenges. IoT forensics is a branch of digital forensics dealing with IoT-related cybercrimes and includes the investigation of connected devices, sensors and the data stored on all possible platforms.

According to the literature, several of the current limitations of IoT forensics include the management of different streams of data sources, the complicated three-tier architecture of IoT, the lack of standardized systems for capturing real-time logs and storing them in a valid uniform form, the preparation of highly detailed reports of all information gathered its corresponding representation, the preservation and acquisition of evidence considering its volatility and value of data, and the adoption of routine forensic tasks in the IoT ecosystem \cite{Hou2020,Stoyanova2020,kamal2021review,atlam2020internet,lutta2021complexity}. Data encryption trends also present additional challenges for IoT forensic investigators, and arguably cryptographically protected storage systems is one of the most significant barriers hindering efficient digital forensic analysis \cite{Sayakkara2019,kamal2021review}. Other studies highlight additional limitations of IoT forensics processes such as interoperability and availability issues related to the vast amount of connected IoT devices \cite{Yakubu2018,kamal2021review,atlam2020internet,lutta2021complexity}, the Big Data nature of the IoT forensics evidence (Variety, Velocity, Volume, Value, Veracity) \cite{Koroniotis2019,atlam2020internet} and the various security storage challenges of IoT forensics evidence, especially when related to biometric data \cite{Ross2020}. Finally, various regulatory-related challenges also exist in the IoT forensics domain, particularly issues relevant to the ownership of data in the cloud as defined by region-specific laws \cite{Yakubu2018,kamal2021review,atlam2020internet,lutta2021complexity}. For instance, service-level agreements stipulating the ``terms of use'' of the cloud resources between the cloud customer and the cloud service provider do not incorporate forensic investigations' provisions. Legislative frameworks adopted in specific regions, such as the GDPR in Europe, also pose significant challenges for IoT forensic investigations, particularly data privacy provisions \cite{Stoyanova2020,kamal2021review,atlam2020internet,lutta2021complexity}. Finally, the use of blockchain and its capability to enhance IoT forensic investigations has been also discussed in \cite{kamal2021review}.  In Table \ref{tab:iotforensics} we provide a classification of the current challenges of IoT forensics approaches.

\begin{table}[hbt!]
 \centering
 \begin{footnotesize}
 \rowcolors{2}{gray!25}{white}
 \resizebox{\linewidth}{!}{
  \begin{tabular}{p{.65\linewidth}>{\raggedright\arraybackslash}p{.25\linewidth}}
\toprule {\textbf{Challenge/Limitation}} & \textbf{References}\\
 \midrule
Heterogeneous data acquisition, integrity and interpretation & \cite{Hou2020,Stoyanova2020,Yakubu2018,Koroniotis2019,kamal2021review,atlam2020internet,lutta2021complexity} \\
Lack of training and interoperability between investigators and court & \cite{Hou2020,Stoyanova2020} \\
Forensic process in IoT environment may necessitate all three levels including device level forensics, network forensics, and cloud forensics & \cite{Hou2020} \\
Lack of forensic readiness mechanisms and their ethical management & \cite{Hou2020,Stoyanova2020,atlam2020internet} \\
Availability of IoT devices due to their resource-constraint nature & \cite{Koroniotis2019,lutta2021complexity} \\
Cross border investigations due to different jurisdictions and laws & \cite{Koroniotis2019,Yakubu2018,kamal2021review,atlam2020internet,lutta2021complexity} \\
Volume of evidence storage and logging-related issues & \cite{Koroniotis2019,Stoyanova2020,kamal2021review,lutta2021complexity} \\
Data encryption mechanisms and cryptographically protected storage systems & \cite{Sayakkara2019,kamal2021review,lutta2021complexity,atlam2020internet} \\
Sound and standardized methodologies, evaluation procedures and benchmarks & \cite{Stoyanova2020,Ross2020,Yakubu2018,kamal2021review,atlam2020internet,lutta2021complexity} \\
Update forensic tools to fight novel cybercrime & \cite{Stoyanova2020,kamal2021review,atlam2020internet} \\
Lack of provision regarding forensic investigations or evidence recovery in service level agreements between service providers and customers& \cite{Stoyanova2020} \\
 \bottomrule
 \end{tabular}
}
 \caption{High level extraction of limitations in IoT forensics.}
 \label{tab:iotforensics}
 \end{footnotesize}
\end{table}%

\subsection{Filesystems, Memory and Data Storage Forensics}
\label{sec:filesystems}

Forensic analysis of large filesystems requires efficient methods to manage the potentially large amount of files and data contained in them. System logs are one of the most used information sources to leverage forensic investigations. In \cite{studiawan_19} the authors provide a review of the publicly available datasets used in operating system log forensics research and taxonomy of the different techniques used in the forensic analysis of operating system logs. The taxonomy is created based on a common investigation format that includes event logs recovery, event correlation, event reconstruction and visualization. 
Distributed filesystem forensics is even a more challenging task, such as in the case of identifying the malicious behaviour of the attackers by analysing cloud logs \cite{khan_16}. Nevertheless, the accessibility attributes associated with cloud logs impede the goals of investigating such information, as well as other challenges, similar to those extracted in Section \ref{sec:cloud_taxonomy}. 

Another challenging area is the analysis of proprietary systems such as SCADA systems. In \cite{awad_18} the authors present a survey on digital forensics that are applied to SCADA systems. The survey describes the challenges that involve applying digital forensics to SCADA systems as well as the range of proposed frameworks and methodologies. The work also focuses on the research that has been carried out to develop forensic solutions and tools that can be tailor-made for the SCADA systems. 
Recent research has revealed that malware developers have been using a broad range of anti-forensic techniques and escape routes in-memory attacks and system subversion, including BIOS and hypervisors. In addition, code-reuse attacks such as returned oriented programming pose a serious remote code execution threat. To neutralise the effects of malicious code, specific techniques and tools such as transparent malware tracers, system-wide debuggers were proposed. In \cite{botacin_18}, authors present a survey on the state-of-the-art techniques that demonstrate the capability of thwarting the anti-forensic strategies previously mentioned.


Memory forensics refers to the forensic analysis of a system’s memory dump. A system's memory can contain evidence related to the usage of the system, including the list of running processes, network connections, or the keys for the driver’s encryption. Usually, such data are not stored in the permanent storage of the system and are completely lost when the system is turned off or unplugged from the power.
In the literature, we can find surveys devoted to the analysis of the memory acquisition techniques \cite{latzo19,osbourne_13} (i.e., both hardware and software-based), the subsequent memory analysis \cite{case_17}, and the available tools \cite{osbourne_13}. The main challenges of memory forensics derive from the fact that memory is volatile, so it has to be acquired when the system is running and thus probably modified by the running applications. This can lead to the page smearing issue \cite{case_17}, i.e., inconsistencies between the state of the memory as described by the page tables compared with the actual contents of the memory. Another issue that can occur during the memory acquisition is the incorporation of pages, which are not present in the memory due to page swapping or demand paging \cite{case_17}. Finally, although the memory acquisition techniques should be OS and hardware agnostic \cite{latzo19}, each OS architecture handles the memory differently and is equipped with distinctive tampering protection mechanisms that hinder access to memory.


A database (DB) is the most traditional way to organise and store data. The majority of applications and online services deploy some type of DB to store records about their customers, financial records, inventory, etc. Besides the vast amount of data that could be contained in a DB, a database management system (DBMS) which allows the end-users to administer the DB and store and access the data in a specific format, can also provide evidence of actions in user-level granularity. For instance, it can reveal who and when stored/accessed specific records. Therefore, digital forensics for DB has attracted the attention of the research community \cite{9110909}. From this perspective, several surveys focused on database digital forensics based on the log files, metadata, and similar types of artefacts for the case of relational and NoSQL DB \cite{adedayo15,chopade19,hauger18}. Furthermore, other authors addressed the digital forensic opportunities on the procedure of data aggregation and analysis, as well as their structural architecture to benefit forensic procedures \cite{9110909,jusas2017methods}. Digital triage is of special relevance here since reviewing many potential sources of digital evidence for specific information by using either manual or automated analysis is mandatory to enhance investigations \cite{jusas2017methods}. Nevertheless, the authors highlight that the legitimacy of several acquisition procedures is constrained by the applicable legislation and that the current state of practice requires more efficient solutions, especially when dealing with huge amounts of data. In \cite{al2021face}, the authors presented a framework for database forensic investigations enhanced by forensic experts' opinions with the aim to overcome the main issues that investigator's face, such as the lack of standardized tools and different data structures and log structures.

Considering the increasing amount of IoT technologies and small devices that require live data analysis due to the volatility of the data stored in them, it is crucial to develop new strategies to enhance data acquisition procedures \cite{sutherland_08}. In the context of database forensics and data acquisition, the challenges of big data analysis and data mining techniques for digital forensics \cite{beebe06,Quick2014}, and text clustering \cite{almaslukh19} were investigated.  Moreover, a survey of techniques to perform similarity digest search is provided in \cite{moia2017similarity}.

Table \ref{tab:database} summarises the main limitations and challenges extracted from the literature analysed in this section.

\begin{table}[hbt!]
 \centering
 \rowcolors{2}{gray!25}{white}
 \begin{footnotesize}
 \resizebox{\linewidth}{!}{
   \begin{tabular}{p{.65\linewidth}>{\raggedright\arraybackslash}p{.25\linewidth}}
\toprule {\textbf{Challenge/Limitation}} & \textbf{References}\\
 \midrule
 Performance issues and logging inducing overhead in terms of query latency, storage, etc. & \cite{adedayo15,moia2017similarity,jusas2017methods, studiawan_19,botacin_18,chopade19}\\ 
Lack of standardized tools and technologies & \cite{adedayo15,almaslukh19,jusas2017methods,khan_16,al2021face,9110909}\\
Forensic seizure and analysis of proprietary and/or distributed filesystems & \cite{studiawan_19,khan_16,awad_18,al2021face,Quick2014,sutherland_08,sutherland_08} \\ 
Variety of format and content type. Not standard logging features and settings & \cite{adedayo15,hauger18,moia2017similarity,jusas2017methods,botacin_18,al2021face,Quick2014,chopade19,9110909}\\
No validation/verification in real-life scenarios and large datasets & \cite{beebe06,almaslukh19} \\
Subjectivity of the evaluation of content retrieval algorithms & \cite{beebe06,almaslukh19} \\
Advanced knowledge and training of analysts and investigators & \cite{beebe06,jusas2017methods} \\
Lack of guidance for investigators regarding selective search and seize. Subjectivity of search terms based on investigator’s experience & \cite{almaslukh19,jusas2017methods} \\
Difficulty to apply low-level analysis techniques, hindering correctness of the results & \cite{latzo19,al2021face} \\
Sophisticated malware implementing anti-forensic techniques & \cite{latzo19,almaslukh19,botacin_18,osbourne_13} \\
Volatile data acquisition due to hardware constraints & \cite{sutherland_08} \\ 
Stealthy non-memory-resident malware & \cite{case_17}\\ 
Handling, execution and monitoring of memory & \cite{ osbourne_13,case_17,botacin_18}\\
Physical access to RAM & \cite{latzo19,sutherland_08}\\
Accurate similarity search of documents and Dynamic insertion/deletion of elements & \cite{moia2017similarity}\\
 \bottomrule
 \end{tabular}
}
 \caption{High level extraction of challenges in file system, memory and data storage forensics.}
 \label{tab:database}%
 \end{footnotesize}

\end{table}%

\subsection{Blockchain}
\label{sec:blockchain}

Blockchain technology has been constantly integrated into existing systems or used as the basis to rebuild systems from scratch in various domains. Besides the financial domain to which it was initially applied, through bitcoin, blockchain technology is currently used in various other use cases such as supply chain management, cybersecurity enhancement, document/certificates validation, crowdfunding campaigns, and more \cite{casino2018systematic}. Additionally, because financial system set on blockchain provide more privacy than traditional payment systems, it is common for cryptocurrencies to be used for criminal activities \cite{jose_17}. This sets blockchain forensics methodologies as a necessity \cite{dasaklis_20} due to the large volume of data that are stored in blockchain systems and the number of processes that are managed by such systems. The main property of blockchain-based systems is the guaranteed protection of data integrity, which is directly related to forensic analysis. On the one side, this property makes forensic analysis more manageable. However, on the other side, this may complicate the process as users may be more cautious when interacting with such systems.

It has to be noted that a large portion of blockchain systems are public, allowing access to everybody and thus making forensic analysis a surplus process. A forensics investigator can set up a node in a public blockchain network, sync it with the rest of the nodes and obtain a local copy of the ledger. Even in such cases, the structure of the information stored in the ledger of blockchain systems is not optimal with respect to retrieving all required data (e.g., for a specific account or a specific smart contract), so efficient mechanisms are required \cite{balaskas_18} to extract valuable information out of the large volume of data stored in public ledgers \cite{turner_18}. In the case of private blockchain systems, the ledger data are not publicly available and traditional forensics approaches have to be applied to blockchain nodes to extract data.

Even if data are by default publicly available, it is still challenging to identify malicious activity on such platforms. It is common for deployed smart contracts to suffer from various vulnerabilities either due to poor implementation or not properly configured blockchain networks \cite{chen_20}. In such cases, users can take advantage of such vulnerabilities, mainly aiming at financial profit. It is challenging to detect such activity and identify the actors that have initiated it. Smart contracts execution is not a straightforward process, and past execution cannot be easily repeated in a forensic sound way \cite{homoliak_20}. Apart from that, smart contracts may also get self-destructed by a special OPCODE that makes following past transactions even harder \cite{wang_20}.

Furthermore, privacy concerns have been raised concerning early open public blockchain systems, and thus, there have been multiple alternative systems that make use of various privacy-enhancing techniques such as zero-knowledge proofs, onion routing or ring confidential transactions to protect users privacy \cite{koerhuis2020forensic}. In such cases, forensics analysis of either network nodes or users’ wallets is required to retrieve either logs or cryptographic keys that can be used along with data existing on public ledgers and provide more information about the transactions that have taken place.

While the data stored in the ledger are of great importance, there are more data to be considered when analyzing a blockchain node. The ledger holds all committed transactions, but a blockchain node stores more information with respect to its interactions with other nodes or clients. For example, the IP of the client that has connected to a node to submit a transaction or the activity of a specific node in the network (e.g., sync requests) are not included in the ledger’s data.
On top of those, multiple security blockchain attacks are mainly targeted against the infrastructure or the network's backbone and not against its content. Mining attacks, network and  long-range attacks \cite{saad_20,8653269} target at taking control of the blocks formation process, to maliciously alter past committed transactions and achieve double-spending attacks. In such cases, digital evidence from deployed nodes is the only way to prove malicious activity. At the same time, the size of the network in public blockchain systems makes it even harder to retrieve the required evidence. Table \ref{tab:blockchain} summarises the main challenges extracted from the blockchain forensics literature.

\begin{table}[t]
 \centering
 \begin{footnotesize}
 \rowcolors{2}{gray!25}{white}
  \begin{tabular}{p{.65\linewidth}>{\raggedright\arraybackslash}p{.25\linewidth}}
\toprule {\textbf{Challenge/Limitation}} & \textbf{References}\\
 \midrule
 Acquisition of large volume of data & \cite{balaskas_18,turner_18} \\ 
 Inefficient data structures and lack of standardized analysis & \cite{balaskas_18} \\ 
 Privacy preserving mechanisms that hinder data acquisition & \cite{koerhuis2020forensic} \\
 Difficulties in exploring smart contracts execution & \cite{chen_20,homoliak_20,saad_20,wang_20} \\
 Mining and network attacks & \cite{saad_20} \\

 \bottomrule
 \end{tabular}
 \end{footnotesize}
  \caption{High level extraction of challenges in blockchain forensics.}
 \label{tab:blockchain}%
\end{table}%

\subsection{Multimedia}
\label{sec:multimedia}
Due to the increasing number of ubiquitous technologies (e.g., IoT devices, smartphones, wearables) leveraged by the 4${^{th}}$ industrial revolution, as well as a substantial improvement in the connectivity capabilities in smart scenarios due to 5G, the amount of multimedia prosumers (i.e., both producers and consumers of data) is increasing dramatically year after year \footnote{\url{https://wearesocial.com/blog/2020/01/digital-2020-3-8-billion-people-use-social-media}, \url{https://www.cisco.com/c/en/us/solutions/collateral/executive-perspectives/annual-internet-report/white-paper-c11-741490.html}}. Nevertheless, such multimedia content growth is a double-edged sword. On the one hand, it is a synonym of opportunities for the industry, companies and users. On the other hand, it augments the possible vulnerabilities and attack vectors of such systems, which malicious users can exploit.

Digital forensics in the context of multimedia has received substantial attention from the research community. There exist numerous image forgery detection surveys exploring the topic from a global perspective \cite{ farid09, daCosta2020, Saber2020361, zheng19, Bourouis20201, Kaur20201281, Ansari2019ACA,pandey2016passive,gupta2021passive}. In this context, pixel-based image forgery detection is one of the main topics \cite{qureshi15}, including image splicing forgery \cite{abrahim17}, and copy-move forgery \cite{dixit17, teerakanok19, zhang18}, which is a well-known technique in which parts of the current images are used to cover/hide specific characteristics. Some authors focused on passive techniques to detect forgery \cite{birajdar13}, or carving on specific file formats such as JPEG \cite{ali18}.
Other image forensics surveys analysed topics such as hyperspectral image \cite{khan18,daCosta2020}, image authentication \cite{korus17}, the affectation of noise in images \cite{julliand16} and image steganalysis \cite{chutani19,karampidis18,luo11,Yang2020,Dalal2020}. Another set of surveys focus on the specific context of child abuse material and its detection through image and video analysis \cite{franqueira_18,sanchez_19,cifuentes2021survey,acar_18}. More recently, the advent of deep learning techniques has enhanced the capabilities of image integrity detection and verification, outperforming traditional methods in several image-related tasks, especially in these where anti-forensic tools were used \cite{Yang2020, Dalal2020,NOWROOZI2021102092}.
In the context of video files, we can find surveys on video steganalysis \cite{dalal18,Dalal2020,Yang2020}, video forgery detection \cite{Kaur20201281,kingra16,Dalal2020,Bourouis20201,Shelke2020,pandey2016passive}, video forensic tools \cite{shahraki13,Bourouis20201,Yang2020,Alsmirat2020437}, video surveillance analysis \cite{becerra-riera19,t19}, and video content authentication \cite{singh18}. Finally, digital audio forensics has also been studied in \cite{zakariah18}. Table \ref{tab:multimedia} summarises the main limitations and challenges extracted from the multimedia forensic literature.

\begin{table}[hbt!]
 \centering
 \begin{footnotesize}
 \rowcolors{2}{gray!25}{white}
   \begin{tabular}{p{.5\linewidth}>{\raggedright\arraybackslash}p{.4\linewidth}}
\toprule {\textbf{Challenge/Limitation}} & \textbf{References}\\
 \midrule
 Standardized evaluation procedures and benchmarks & \cite{zhang18,birajdar13,qureshi15,korus17,zheng19,singh18,kingra16,dalal18,franqueira_18,sanchez_19,Yang2020,Dalal2020,cifuentes2021survey,acar_18,dixit17,pandey2016passive,gupta2021passive} \\
 Explore the use of novel AI methods and novel data types & \cite{zhang18,zakariah18,becerra-riera19, korus17,chutani19,karampidis18,teerakanok19,singh18,sanchez_19,Yang2020,Dalal2020,Bourouis20201,Shelke2020,daCosta2020,NOWROOZI2021102092,Ansari2019ACA,cifuentes2021survey,t19,khan18,gupta2021passive} \\
 Robust pre-processing and feature extraction & \cite{zakariah18,becerra-riera19,abrahim17,birajdar13,julliand16,chutani19,zheng19,ali18,teerakanok19,singh18,kingra16,Yang2020,Bourouis20201,Shelke2020,NOWROOZI2021102092,pandey2016passive} \\
 Reduce training and data acquisition overheads & \cite{becerra-riera19,chutani19,shahraki13,franqueira_18,sanchez_19,Saber2020361,Dalal2020,Bourouis20201,Shelke2020,Ansari2019ACA,acar_18,gupta2021passive} \\
 More comprehensive outcome readability & \cite{birajdar13,zheng19,teerakanok19,dalal18,franqueira_18,NOWROOZI2021102092,Ansari2019ACA} \\
 Increased effort to circumventing anti-forensic techniques & \cite{birajdar13,farid09,qureshi15,zheng19,singh18,franqueira_18,Yang2020,Bourouis20201,Alsmirat2020437,Shelke2020,NOWROOZI2021102092} \\
 Rigorous mechanisms to ensure protection/watermarking & \cite{qureshi15,korus17} \\
 Analyse multiple threats/tampering at once & \cite{qureshi15,karampidis18,zheng19,singh18,shahraki13,dalal18,sanchez_19,Dalal2020,Bourouis20201,Kaur20201281,cifuentes2021survey,luo11,gupta2021passive}\\
 Reliable detection with real data and dynamic contexts& \cite{Dalal2020,Bourouis20201,Alsmirat2020437} \\
 \bottomrule
 \end{tabular}
 \end{footnotesize}
 \caption{High level extraction of challenges in multimedia digital forensics.}
 \label{tab:multimedia}%

\end{table}%

\subsection{Miscellaneous}
\label{sec:miscellaneous}

This section is devoted to the digital forensics reviews that fall beyond the domain categorisation of the previous paragraphs.

As observed in most topics, anti-forensics can be understood as a standalone concern in digital forensics, which requires investigation in each context. The term anti-forensics refers to methods and strategies that prevent forensic investigators and their tools from achieving their goals. There are several examples of anti-forensic methodologies \cite{CONLAN2016S66}, such as encryption, data obfuscation (e.g., trail obfuscation), artifact wiping, steganography and image tampering \cite{8853452}, protected/hidden communications (e.g., tunnelling, onion routing), malware anti-sandbox/debug, VM and in general anti-analysis methods \cite{244694,chen2016advanced,bulazel2017survey,branco2012scientific}, and spoofing. As stated in \cite{HARRIS200644}, anti-forensic methods exploit the dependence of human elements on forensic tools, as well as the limitations of the underlying hardware in terms of architecture and computational power. Therefore, enhancing the training and knowledge level of investigators and more robust forensic procedures (e.g., anti-anti forensic techniques \cite{8853452}) are critical to minimise the impact of anti-forensics. In this line, some authors argue that the use of proactive forensics models could help enhancing the robustness of forensic investigations \cite{Alharbi2011a}.

Another emerging topic in digital forensics is related to unmanned aerial vehicles (UAVs), or more commonly known as drones \cite{al2021research}. The applications and versatility of these devices are becoming more popular in a myriad of contexts, from industrial to military applications. One of the main challenges of drone forensics is the set of different hardware components that are part of a drone \cite{horsman2016unmanned}, and the particular treatment that they require (i.e., with special regard to advanced anti-forensic techniques taking place \cite{Atkinson2021}, as well as the necessity of live forensics \cite{adelstein2006live,al2021research} in this context). For instance, drones consist of sensors, flight controllers, electronic and hardware components, on-board computers, and radiofrequency receivers, each one linked to one or many evidence sources in terms of, e.g., data storage (the different memory sources present in the drone, such as memory cards storing media, or other software), data communications and other logs and data stored in sources related to the drone, such as the drone controller and external cloud-based sources \cite{RENDUCHINTALA201952,patsakisdrones}. At the moment of writing, there are no baseline principles, standards, nor legislation covering all the particularities of forensic drone investigations \cite{patsakisdrones,al2021research}. Thus, efforts towards the establishment of sound protocols, specific forensic frameworks, as well as drone-based forensic tools are critical \cite{al2021research}.


In \cite{keyvanpour2014digital}, authors surveyed the different dimensions and concerns which digital forensics should cover in the context of social networks. The authors discussed several aspects of social networks, such as privacy and security issues, the criminal and illegal acts that can occur, and the attacks on the underlying platform and the users. In addition, they describe several strategies to detect such abnormal behaviours along with the necessity to develop both pro-active and reactive mechanisms. In terms of community detection, graph analytic methods and tools are crucial to detect criminal networks in different contexts, such as finance, terrorism, and other heterogeneous sources \cite{sangkaran_20}. In \cite{pasquini2021media}, authors surveyed the efforts done so far on the analysis of social network shared data according to source identification, integrity verification and platform provenance. Moreover, authors discussed the current methodologies, and highlighted the current challenges along with the need for multidisciplinary approaches to overcome them.

A sector that is receiving increasing attention due to its critical relevance to the proper functioning of our society is the energy sector, and more concretely, the smart grid. In \cite{parra2019implementation}, authors explore practical cybersecurity models and propose some guidelines to enhance the protection of the smart grid against cyber threats. Moreover, they explore software-defined networks and their main benefits and challenges. Finally, the authors propose a conceptual forensic-driven security monitoring framework and highlight the relevance of forensics by design in development phases. Context-aware scenarios such as smart cities have been also receiving increased attention due to their complex structures, requiring the continuous data collection, processing and interaction between a myriad of devices \cite{batista2021sensors,ahmadi2020cyber}. Digital forensics in this particular scenario is a recent paradigm which requires further efforts from the research community to enhance cyber resilience and to provide efficient incident response mechanisms \cite{ahmadi2020cyber}.

\subsection{Challenge Analysis and Aggregated Results}
\label{sec:aggregated}

The classification of challenges and limitations according to each topic of the taxonomy has been conducted to keep a balance between accurate descriptions of challenges and hierarchical classification. On the one hand, we want to facilitate identifying the gaps and limitations of each topic and provide a clear path for both new and experienced investigators towards the corresponding literature. On the other hand, and as stated in Section \ref{sec:motivation}, we provide the reader with a clear overview of the research lines that should be strengthened in the digital forensics ecosystem, as well as their interrelations according to each topic of our taxonomy. Therefore, we used the extracted challenges of each topic and merged the ones appearing more than once (i.e., the ones appearing only in their corresponding topic were ignored due to their specificity) to create a comprehensive overview of the digital forensics challenges in Table \ref{tab:cross_limitations}. As it can be observed, we identified several limitations of digital forensics that can be applied in several topics or contexts and thus, indicate the need to devote more research efforts towards them. Note that, for instance, the last topic of the Table \ref{tab:cross_limitations} appears to be only affecting IoT, yet we identified this challenge in the miscellaneous topic, and thus, we decided to include it. Nevertheless, since several topics are analysed in such a category, we did not represent them in Table \ref{tab:cross_limitations}.

\begin{table*}[hbt!]
\renewcommand{\tabcolsep}{1mm}
 \centering
 \footnotesize
 \rowcolors{2}{gray!25}{white}
 \resizebox{\linewidth}{!}{
\begin{tabular}{lccccccc}
\toprule
\textbf{Challenge/Limitation} & \textbf{Cloud} & \textbf{Networks} & \textbf{Mobile} & \textbf{IoT} & \textbf{FS \& DB} & \textbf{Blockchain} & \textbf{Multimedia} \\
 \midrule
 Sound data acquisition from heterogeneous/ubiquitous sources& \checkmark & \checkmark & \checkmark &\checkmark &\checkmark & & \checkmark \\
 Anti-forensics and protected storage systems & \checkmark & & \checkmark & \checkmark & \checkmark& \checkmark &\checkmark \\
 Sound and standardized evaluation procedures and benchmarks& & & & \checkmark & \checkmark & \checkmark & \checkmark \\ 
 Lack of jurisdictional and legal requirements for different investigation scenarios &\checkmark & & \checkmark & \checkmark & & & \\ 
 Lack of forensic readiness mechanisms and their management & \checkmark & & &\checkmark & & &\checkmark  \\
 Update forensic tools to fight novel cybercrime & \checkmark& & \checkmark & \checkmark & & & \checkmark  \\ 
 Lack of training and interoperability between investigators and court &\checkmark & & &\checkmark &\checkmark & &  \\ 
 Reduce pre-processing, training and data acquisition overheads & & \checkmark & \checkmark & & & \checkmark &\checkmark  \\
 Cross border investigations due to different jurisdictions and laws & \checkmark & & &\checkmark & & &  \\ 
 Lack of device-based standards and procedural guidelines& & &\checkmark & & \checkmark & & \\ 
 Reliable detection of threats/attacks and testing in real scenarios & & \checkmark & & & \checkmark & & \checkmark  \\ 
 The nature of the devices requires the adoption of live forensics approaches & & & \checkmark &\checkmark & \checkmark& &  \\ 
 Evidence storage and logging-related issues& & & & \checkmark & \checkmark & &  \\ 
 Multiple forensic contexts due to different data sources & & & & \checkmark & & & \\ 
 Lack of trust and robust chain of custody preservation &\checkmark & & \checkmark & & & &  \\ 
 Availability of devices due to their resource-constraint nature& & & & \checkmark & & &  \\ 
 \bottomrule
\end{tabular}
}
\caption{Cross-domain abstraction of the challenges and limitations of digital forensics, ordered by relevance according to the amount of times they were found in the topics of the taxonomy. For the sake of fairness, the general column \textit{Miscellaneous} has been omitted.}
 \label{tab:cross_limitations}%

\end{table*}

The most reported challenge is the sound data acquisition from heterogeneous sources and its interpretation, including different hardware and monitoring processes collecting data and logs dynamically. Note that data acquisition and management is a challenge affecting activities related to digital forensics. Moreover, data fragmentation, a common scenario nowadays, hinders investigations further. It is important to note that data acquisition is critical to creating benchmarks, which help researchers and practitioners to evaluate their models. The latter enables characteristics such as reproducibility and pushes the advancement in the state of the art, which is needed to keep up with the pace of technology development\cite{GARFINKEL2009S2,GRAJEDA2017S94}.
The next most challenging issue is related to anti-forensics methods, which has been discussed in several sections of the taxonomy as well as in Section \ref{sec:miscellaneous}. Anti-forensic strategies leveraged by malicious actors include adversarial methods such as obfuscation or encryption applied to, e.g., data and storage systems, as well as hardware-related technological challenges, such as mobile phones due to their inherent security measures, or in the case of drones due to their specific particularities, and software, as well as in the case of malware. In the case of tools and evaluation benchmarks, it is evident that the community needs to devote more efforts towards fighting novel cybercrime, especially in topics where, e.g., different data sources and technologies are present. For instance, in the case of IoT and UAVs, different data sources may necessitate different digital forensics strategies, including tools related to device level forensics, network forensics, and cloud forensics. Another challenge that affects digital forensics is the lack of jurisdictional and legal requirements for different investigation scenarios such as ethics and data management of confidential and personal data. This is particularly relevant nowadays due to the widespread use of distributed systems such as blockchain and the cloud. The latter means that software and data may reside in different countries, and thus, specific cross-border collaborations are required, adding another layer of complexity to digital investigations. Moreover, this scenario impedes the adoption of proactive measures due to the difficulty of applying measures that conform to different legal frameworks.

 A proper understanding between all the actors involved in the digital forensics context, including stakeholders, LEAs, and court members, is mandatory to ensure the successful prosecution of perpetrators. In this regard, one of the highlighted challenges is to ensure that all partners have a sufficient level of training (including technical knowledge and standardised guidelines) and a proper understanding, including readable reports to enable a fruitful collaboration. Moreover, while it seems procedural, the chain of custody is still a challenge. This can be attributed to multiple reasons, such as obvious negligence of the corresponding personnel to properly report evidence acquisition and/or handling, corrupted officers, or even gaps in the process. Nevertheless, all of them cause severe issues in a court as a case can be missed or misjudged. Secure and auditable means of storing and processing the chain of custody, as proposed by LOCARD\footnote{\url{https://locard.eu/}} with the use of blockchain technology seems like a logical and stable solution. A more thorough description of forensic readability and its challenges is discussed later in Section \ref{sec:forensic_readability}.

 Data acquisition, as previously stated, is not only a challenge in terms of the existing heterogeneous data sources and context but also in terms of size. The big data era comes with a myriad of opportunities but also with their corresponding challenges, since logging and data acquisition in specific scenarios may pose technical challenges. This issue is exacerbated when coupled with cross-border investigation requirements due to data fragmentation. Moreover, once data corresponds to multiple forensic contexts, the complexity of performing digital investigation grows exponentially, leaving aside the need to perform live forensics according to the particularities of the hardware. Additionally, the availability of some devices due to their resource-constraint nature is a further challenge. For instance, IoT botnets have high volatility, and UAVs may implement self-defence mechanisms, even at the physical level. In the case of the \textit{Miscellaneous} category, we included the challenges and limitations of anti-forensics, drone forensics, smart grid, smart cities and social networks.

 According to the outcomes depicted in Table \ref{tab:cross_limitations}, we can observe that topics such as IoT, cloud, and mobile are affected by the highest amount of challenges. Therefore, we believe that researchers and practitioners should devote more efforts to solving such topics' challenges by leveraging cross-domain collaborations to enhance the quality and applicability of their outcomes. Similarly, other challenges which appear in several topics could be tackled more quickly if they were targeted with a multidisciplinary approach, with experts from the different digital forensics topics.

\begin{figure*}[t]
\centering
\includegraphics[trim={0 3.5cm 10cm 0cm},clip,width=\textwidth]{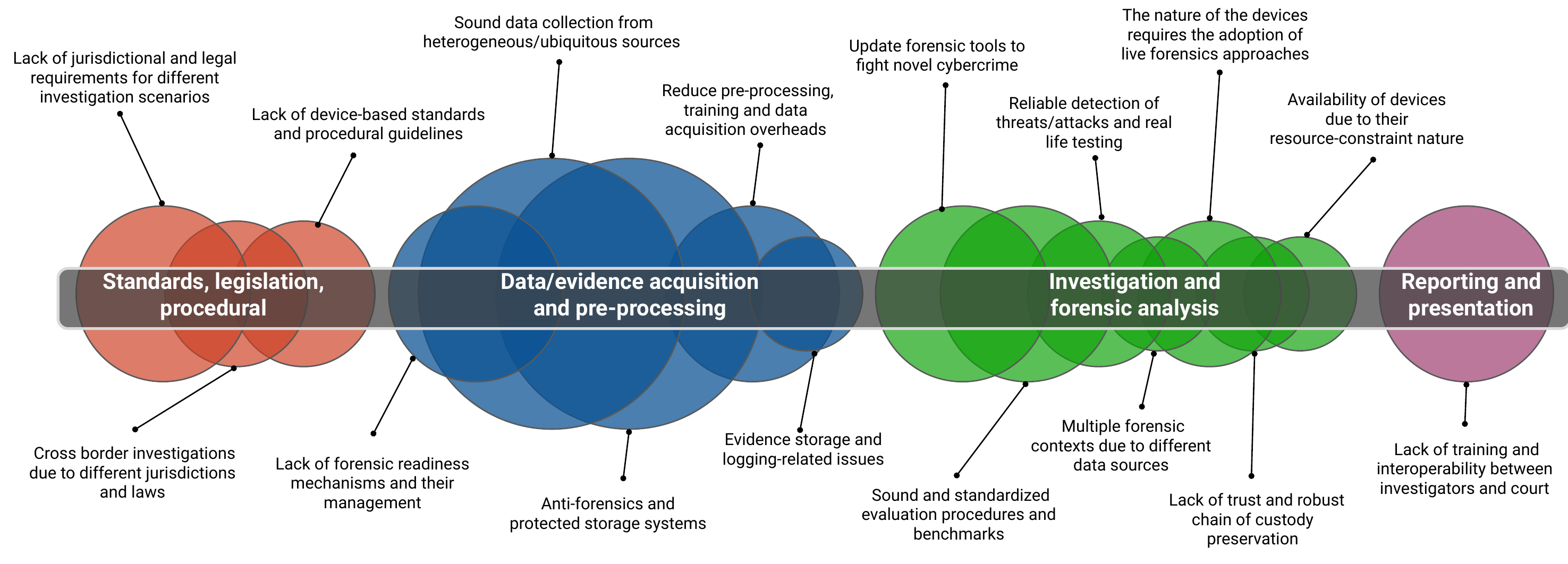}
\caption{Main digital forensic challenges mapped into different categories according to their application context, from the initial steps of an investigation (left) to the final ones (right). The size of each circle denotes the times it appeared considering the topics of the taxonomy.}
\label{fig:circles}
\end{figure*}

 To create a visual representation of these challenges, we believe that mapping each challenge into different categories will highlight which need to be reinforced. Therefore, Figure \ref{fig:circles} presents the outcomes of our taxonomy in terms of topic challenges mapped into different categories representing different phases, from the creation of the legal basis and framework of an investigation to the final reporting of the outcomes. As it can be observed, the challenges most cited in the literature are present in the evidence acquisition and data pre-processing category. They are mainly related to data acquisition issues and anti-forensics. Notably, these challenges affect the forensic procedures from the beginning (i.e., if we do not consider the standards, legislation and procedural category), and thus, it is crucial to devote efforts to overcome them.
The investigation and forensic analysis category contains the highest number of challenges. Therefore, the topics identified in the taxonomy share similar technical concerns in their corresponding contexts, and more multidisciplinary collaboration is needed towards such direction. The reporting and presentation category highlights one yet critical challenge since the proper reporting of an investigation affects the outcome of the whole investigation. We further discuss about forensic readability and reporting in Section \ref{sec:forensic_readability}.

\section{Digital Forensics Methodologies, Practices and Standards}
\label{sec:digitalforensics_mps}

In addition to the topic-based taxonomy presented in Section \ref{sec:taxonomy}, we collected a set of literature reviews, included in our research methodology, that analysed forensic frameworks and process models, forensic tools, and the adaptability and forensic readiness of the actual practices. In the following sections, we analyse the content of such reviews by extracting the challenges and identifying the main qualitative features required to achieve forensically sound investigations.

\subsection{Forensic Frameworks and Process Models}

A digital forensics framework, also known as a digital forensics process model, is a sequence of steps that, along with the corresponding inputs, outputs and requirements, aim to support a successful forensics investigation \cite{kohn2006framework,halboob2012state}. A digital forensics framework is used by forensics investigators and other related users to ease investigations and the identification and prosecution of perpetrators. In addition to a set of specific steps identifying each investigation phase, the use of digital forensic frameworks enables timely investigations, as well as a proper reconstruction of the timeline of events and their associated data. In this regard, one of the most critical aspects of a digital investigation is the proper preservation of the evidence chain of custody, since it could lead to unsolvable inconsistencies, risking the admissibility of evidence in court.

According to their phases and their granularity, there are different investigation models suitable for different types of investigations. In this regard, Kohn et al. provide \cite{kohn2013integrated} an integrated suitability framework that maps a set of requirements derived from an ongoing investigation to the most suitable forensic procedure. Moreover, the authors also use a graph-based approach to associate the most well-known forensic frameworks and their interrelationships regarding the number of phases and their content. Other well-known frameworks include the Analytical Crime Scene Procedure Model (ACSPM) \cite{BULBUL2013244}, the Systematic digital forensic investigation model (SRDFIM) \cite{agarwal2011systematic}, and the advanced data acquisition model (ADAM) \cite{adams2013advanced}. In general, law enforcement agencies follow variants of the ACPO (Association of Chief Police Officers) guidelines \cite{williams2012acpo}. Finally, other forensic guidelines and models proposed by NIST and INTERPOL can be found in \cite{kent2006sp, interpolguidelines19}. The most well-known digital forensic frameworks are summarised in Table \ref{tab:mainmodels}.


\begin{table*}[ht]
 \centering
\footnotesize
 \begin{tabular}{p{.7\textwidth}cc}
 \toprule
 \textbf{Name} & \textbf{Year} & \textbf{Reference} \\
 \midrule
Digital Forensic Investigation Model & 2001 & \cite{kruse2001computer}\\
Digital Investigative Process Model & 2001 & \cite{palmer2001road}\\
Abstract Digital Forensic Model & 2002 & \cite{reith2002examination}\\
Integrated Digital Investigation Model & 2004 & \cite{carrier2003getting}\\
Enhanced Digital Investigation Process Model& 2004 & \cite{baryamureeba2004enhanced}\\
Extended Model of Cybercrime Investigation & 2004 & \cite{ciardhuain2004extended} \\
NIST Guide to Integrating Forensic Techniques into Incident Response & 2006 & \cite{kent2006sp}\\
Digital Forensic Model for Digital Forensic Investigation &2011 & \cite{ademu2011new}\\
Systematic digital forensic investigation model & 2011 & \cite{agarwal2011systematic} \\
ACPO guidelines & 2012 & \cite{williams2012acpo}\\
Analytical Crime Scene Procedure Model & 2013 & \cite{BULBUL2013244} \\
Advanced data acquisition model & 2013 & \cite{adams2013advanced} \\
INTERPOL Guidelines for Digital Forensics Laboratories & 2019 & \cite{interpolguidelines19}\\
ENFSI Guidelines & 2016-2020 & \cite{enfsiguidelines} \\
 \bottomrule
 \end{tabular}
  \caption{Most well-known forensic models and guidelines.}
 \label{tab:mainmodels}
 \end{table*}

In general, the procedures summarised in Table \ref{tab:mainmodels} have a common hierarchical structure \cite{yusoff2011common,forensic_guidelines}, which can be divided in the steps described in Table \ref{tab:forensicsteps}. Note that some of the models may include more granular approaches to some of the steps, which are necessary due to the investigation context (e.g., specific devices and seizure/acquisition constraints).

\begin{table}[hbt!]
\rowcolors{2}{gray!25}{white}
 \centering

 \scriptsize
 \setlength{\tabcolsep}{5pt}
 \begin{tabular}{lp{.65\linewidth}}
 \hline
 \textbf{Forensic Step} & \textbf{Description} \\
 \hline
Identification & Assess the purpose and context of the investigation. Initialize and allocate the resources required for the investigation, such as policies, procedures and personnel. \\
Collection \& Acquisition & The seizure, storage and preservation of digital evidence. Although this two steps need to be strictly differentiated in the physical forensics context, a more relaxed approach can be considered in the digital context. \\
Analysis & The identification of tools and methods to process the evidence and the analysis of the outcomes obtained \\
Reporting \& Discovery & The proper presentation of the reports and information obtained during the investigation to be disclosed or shared with the corresponding entities, including the court. \\
Disposal & The relevant evidence are either properly stored for future references or erased. In specific cases, evidence are returned to the corresponding owners. \\
 \bottomrule
 \end{tabular}
  \caption{Main steps in a digital forensic investigation model.}
 \label{tab:forensicsteps}
\end{table}

In the case of the chain of custody and trail of events preservation, a forensically sound procedure needs to ensure features such as integrity, traceability, authentication, verifiability and security \cite{Bonomi2020,Tian2019}. In this regard, Table \ref{tab:chainofcustody} provides a description of each feature.

\begin{table}[hbt!]
\rowcolors{2}{gray!25}{white}
 \centering

 \scriptsize
 \setlength{\tabcolsep}{5pt}
 \begin{tabular}{lp{.75\linewidth}}
 \hline
 \textbf{Feature} & \textbf{Description} \\
 \hline
Integrity & The events data as well as evidences cannot be altered or corrupted during the transferring and during analysis. \\
Traceability & The events and evidence can be traced from their creation till their destruction. \\
Authentication & All the actors and entities are unique and provide irrefutable proof of identity. \\
Verifiability & The transactions and interactions can be verified by the corresponding actors. \\
Security & Only actors with clearance can add content to an investigation or access to it. \\
 \bottomrule
 \end{tabular}
  \caption{Main features required to guarantee chain of custody preservation.}
 \label{tab:chainofcustody}
\end{table}


In the past, several authors identified several challenges in digital investigation processes \cite{greenfield2002cyber,reilly2010cloud,GARFINKEL2010S64,guarino2013digital,mohay2005technical,Quick2014,li2021threat,9160916}, mainly related to the chain of custody preservation, the growth of the data to be processed, and privacy and ethical issues when collecting such data. In addition, our research methodology identified several literature reviews which discussed the challenges and limitations of forensic frameworks. For instance, in \cite{abulaish_18}, the authors leveraged a summary of digital forensic frameworks and tools as well as their interrelationships by using a graph analysis methodology. In addition, they discussed some challenges and limitations of privacy-preserving digital investigation models and proposed some measures to palliate them. In \cite{agarwal_15} the authors presented a chronological review of the most well-known forensic frameworks and their characteristics. The work presented in \cite{amann_15} evaluates the current frameworks among European law enforcement agencies, identifying and defining elements of robustness and resilience in the context of sustainable digital investigation capacity so that organisations can adapt and overcome deviations and novel trends. In \cite{9160916}, the authors identified the need to define specific models according to the forensic context, such as in the case of Mobile Forensics \cite{9160916}. Moreover, the authors proposed a specific forensic framework to improve Mobile Forensics investigations. Further reviews of the most used forensic frameworks and their features can be found in \cite{montasari_16,sabillon_17}.
Table \ref{tab:frameworks_challenges} reports the main challenges in forensic frameworks identified by each literature review.

\begin{table}[hbt!]
 \centering
 \begin{footnotesize}
 \rowcolors{2}{gray!25}{white}
  \begin{tabular}{p{.6\linewidth}>{\raggedright\arraybackslash}p{.3\linewidth}}
\toprule {\textbf{Challenge/Limitation}} & \textbf{References}\\
 \midrule
 Privacy and ethical data management & \cite{abulaish_18} \\
 Seize and investigate big volumes of data & \cite{abulaish_18,amann_15,li2021threat,9160916} \\
 Cross-border models and chain of custody preservation& \cite{ abulaish_18,amann_15,montasari_16,sabillon_17} \\
 Adaptable frameworks for novel cybercrime campaigns& \cite{amann_15,sabillon_17,agarwal_15,9160916} \\
 Effective reporting readability and complexity & \cite{amann_15,li2021threat} \\
 Training and collaboration between stakeholders involved in forensic investigation and prosecution & \cite{amann_15} \\
 Cross-domain technical challenges, technologies, anti-forensics & \cite{ montasari_16,sabillon_17,9160916} \\
 \bottomrule
 \end{tabular}
 \end{footnotesize}
  \caption{High level extraction of challenges reported in forensic frameworks literature reviews.}
 \label{tab:frameworks_challenges}%
\end{table}%

In parallel to forensic guidelines and frameworks, standards are crucial to ensure conformance and mutual compliance across geographical and jurisdictional borders. There are currently numerous standards and established practices provided by organisations worldwide using accepted methods. The technical details on how to forensically approach a given investigation differ depending on the device. The analysis of electronic evidence is typically categorised into the phases stated in Table \ref{tab:forensicsteps}. However, the exact phases naming may vary due to different forensic models' usage according to each organisation's needs.

\begin{figure*}[hbt!]
\centering
 \includegraphics[width=.66\textwidth]{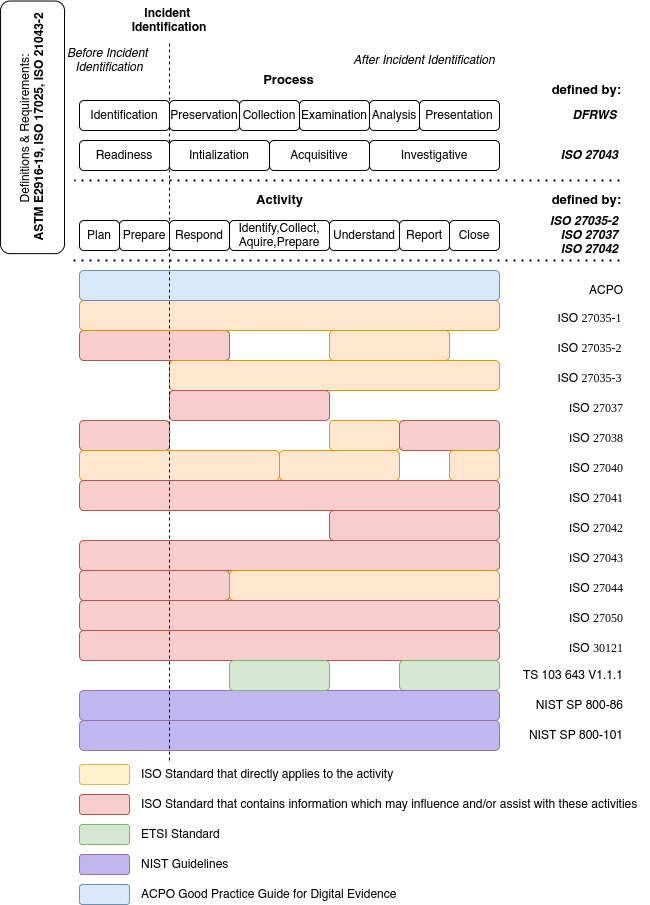}

 \caption{Applicability of standards and guidelines to the investigation process classes and activities.}
 \label{fig:ISO-DF}

\end{figure*}

While not an official standard, the Cyber-investigation Analysis Standard Expression (CASE)\footnote{\url{https://caseontology.org/}} is a community-driven standard that aims to develop an ontology that can efficiently represent all exchanged information and roles within the context of investigations regarding digital evidence. The International Organization for Standardization (ISO) has released a series of standards to assist in this effort by providing the family of ISO 27000, focusing on information security standardisation procedures. In what follows, we present the most relevant standards about digital forensics investigations, which are summarised in Figure \ref{fig:ISO-DF}.

\begin{description}
\item \textbf{ISO/IEC 17025:2017: } In some terms, this standard can be considered an ``infrastructure'' standard for forensic labs. It defines the managerial and technical requirements that testing and calibration laboratories must conform to to ensure technical competence and guarantee that their test are calibration results are acceptable by the corresponding suppliers and regulatory authorities.

\item \textbf{ASTM E2916-19:} The goal of this standard is to assemble the necessary technical, scientific and legal terms and the corresponding definitions in the context of the examination of digital and multimedia evidence. Therefore, the standard spans to various areas such as computer forensics, image, audio and video analysis, as well as facial identification. As a result, ASTM E2916-19 creates a common language framework for all.

\item \textbf{ISO 21043-2:2018:} This standard specifies many requirements for the forensic processes in focusing on recognition, recording, collection, transport and storage of items of potential forensic value. It includes requirements for the assessment and examination of scenes but is also applicable to activities that occur within the facility. This document also includes quality requirements.

\item \textbf{ISO/IEC 27035:} This is a three-part standard that provides organisations with a structured and planned approach to the management of security incident management covering a range of incident response phases

\item \textbf{ISO/IEC 27037:2012:} This standard provides general guidelines about the handling of the evidence of the most common digital devices and \textit{the circumstances including devices that exist in various forms}, giving the example of an automotive system \cite{iso_27037}.

\item \textbf{ISO/IEC 27038:2014:} Describes the digital redaction of information that must not be disclosed, taking extreme care to ensure that removed information is permanently unrecoverable.

\item \textbf{ISO/IEC 27040:2015:} Provides detailed technical guidance on how organisations can define an appropriate level of risk mitigation by employing a well-proven and consistent approach to the planning, design, documentation, and implementation of data storage security.

\item \textbf{ISO/IEC 27041:2015:} Describes other standards and documents to provide guidance, setting the fundamental principles ensuring that tools, techniques and methods, appropriately selected for the investigation.

\item \textbf{ISO/IEC 27042:2015:} This standard describes how methods and processes to be used during an investigation can be designed and implemented to allow correct evaluation of potential digital evidence, interpretation of digital evidence, and effective reporting of findings.

\item \textbf{ISO/IEC 27043:2015:} It defines the key common principles and processes underlying the investigation of incidents and provides a framework model for all stages of investigations.

\item \textbf{ISO/IEC 27050:} This recently revised standard guides non-technical and technical personnel to handle evidence on electronically stored information (ESI).


\item \textbf{ISO/IEC 30121:2015:} Provides a framework for organizations to strategically prepare for a digital investigation before an incident occurs, to maximise the effectiveness of the investigation.

\end{description}

ETSI is a European Standards Organization that produces standards for ICT systems and services used worldwide, collaborating with numerous organisations. In 2020, ETSI published TS 103 643 V1.1.1 (2020-01) \cite{etsi_2020}, a set of techniques for assurance of digital material in a legal proceeding, to provide a set of tools to assist the legitimate presentation of digital evidence\footnote{\url{https://www.swgde.org/documents/published}}. In the meantime, the National Institute of Standards and Technology (NIST) has released guidelines for organisations to \textit{develop forensic capability} (see also Table \ref{tab:mainmodels}), based on the principles of forensic science in the aspect of the application of science to the law. Still, it should not be used on digital forensic investigations due to subjection to different laws and regulations, as clearly stated in their manual. The scope of NIST guidelines is \textit{incorporating forensics into the information system life cycle} of an organisation. The most relevant guidelines are 800-86 \cite{10.5555/2206298} for Integrating Forensic Techniques into Incident Response and 800-101 \cite{ayers_brothers_jansen_2014} for Mobile Device Forensics.

The Scientific Working Group on Digital Evidence (SWGDE) is an organisation engaged in the field of digital and multimedia evidence to \textit{foster communication and cooperation as well as to ensure quality and consistency within the forensic community}. SWGDE has released several documents to provide the current best practices on a large variety of state of the art forensics subjects. Nonetheless, none of them is targeting or addressing drone forensics's particularities. Finally, a review of the international development of forensic standards can be found in \cite{WILSONWILDE20181}.

\subsection{Forensic readiness}

In the past, forensic investigations leveraged a post-event approach, mainly focusing on the analysis of data related to a past incident. In this regard, forensic readiness in terms of pro-active techniques and protocols appeared to minimise the cost and the impact of incidents and are widely used nowadays \cite{robinson2015digital,tan2001forensic,ariffin2021indicators,reddy_13}.

We can find different research approaches, such as the review conducted in \cite{elyas_15}, in which authors discussed how to achieve forensic readiness by collecting the opinion of experts to elaborate a readiness framework with which improve forensic investigations from an organizational perspective. In the case of \cite{endicott-popovsky_15}, authors discussed forensic readiness and several procedures to achieve it, such as fostering the use of Trusted Platform Modules (TPM). Other authors reviewed measures to achieve forensic readiness in a holistic way \cite{mouhtaropoulos_14,marshall_18,harichandran_16,ozel_18,ariffin2021indicators}, as well as recalling the relevance to include and expand the actual guidelines towards incident response readiness (e.g., as in the drafts of the ISO/IEC JTC 1/SC 27 working groups, and the ISO/IEC 27035), training and collaboration between stakeholders involved in forensic investigations and prosecution, and effective reporting readability and complexity. Table \ref{tab:readiness_challenges} describes the main forensic readiness challenges identified by the authors in the literature.

\begin{table}[hbt!]
 \centering

 \rowcolors{2}{gray!25}{white}
 \begin{footnotesize}
 \begin{tabular}{p{.6\linewidth}>{\raggedright\arraybackslash}p{.3\linewidth}}
\toprule {\textbf{Challenge/Limitation}} & \textbf{References}\\
 \midrule
 Privacy and ethical data management from heterogeneous sources & \cite{park_18,ariffin2021indicators} \\
 Cross-border models and interoperability& \cite{elyas_15,endicott-popovsky_15,park_18,harichandran_16,ozel_18,ariffin2021indicators,mouhtaropoulos_14} \\
 Effective reporting readability and complexity & \cite{elyas_15,endicott-popovsky_15,harichandran_16,marshall_18} \\
 Training and collaboration between stakeholders involved in forensic investigation and prosecution & \cite{endicott-popovsky_15,park_18,ozel_18,ariffin2021indicators,mouhtaropoulos_14} \\
 Cross-domain technical challenges, technologies, anti-forensics & \cite{ariffin2021indicators} \\
 \bottomrule
 \end{tabular}
 \end{footnotesize}
  \caption{High level extraction of challenges reported in forensic readiness literature reviews.}
 \label{tab:readiness_challenges}
\end{table}%

Finally, in Table \ref{tab:frameworks_global_challenges} we provide a qualitative summary of the literature reviewed in \ref{sec:digitalforensics_mps} according to the topics discussed in each article.
From Table \ref{tab:frameworks_global_challenges} we can see that topics such as privacy and ethics and the suitability of frameworks that are being proposed to fight novel cybercrime need to be further discussed in the literature. Nevertheless, as previously stated in the article, one of the main challenges is that cybercrime evolves faster than countermeasures and legislations, and thus, investigators are always one step behind.

\begin{table*}[hbt!]
 \centering
 \begin{footnotesize}
 \rowcolors{2}{gray!25}{white}
 \begin{tabular}{cccccccc}
\toprule  &  & \multicolumn{6}{c}{\textbf{Qualitative topic discussion}}\\
 \textbf{Reference}& \textbf{Year} & Frameworks & Privacy/Ethics & Tools & Challenges & Suitability & Adaptability/Readiness \\
 \midrule
 \cite{abulaish_18} & 2018 & \checkmark & \checkmark & \checkmark & \checkmark& \checkmark & \\
 \cite{agarwal_15} & 2015& \checkmark & & & $\circ$ & & \\
 \cite{amann_15} & 2015 & & & & \checkmark & & \checkmark \\
 \cite{arshad_18} & 2018 & & & & \checkmark & & \checkmark \\
 \cite{butler_16} & 2016 & \checkmark & & & \checkmark & $\circ$ & \\
 \cite{9160916} & 2020 & \checkmark & & & \checkmark & \checkmark & \\
 \cite{elyas_15} & 2015 & & & & \checkmark &$\circ$ & \checkmark \\
 \cite{endicott-popovsky_15} & 2015 & & & & $\circ$ & &\checkmark\\
 \cite{harichandran_16} &2016 & & & & \checkmark & & \\
 \cite{montasari_16} & 2016 & \checkmark & & & \checkmark & \checkmark & \\
 \cite{park_18} & 2018 & & \checkmark & & \checkmark & & \checkmark \\
 \cite{sabillon_17} & 2017 & \checkmark & & \checkmark &$\circ$ & & \\
 \cite{li2021threat} & 2021 & &  &  & \checkmark& & \\
 \cite{ariffin2021indicators} & 2021 &  &\checkmark & \checkmark &\checkmark & &\checkmark \\
 \cite{mouhtaropoulos_14} & 2014 &  &\checkmark  & &\checkmark  & & \checkmark \\
 \cite{marshall_18} & 2018 &  &  & $\circ$&\checkmark  & & \checkmark \\
 \cite{ozel_18} & 2018 &  &  & &\checkmark  & & $\circ$\\
 \bottomrule
 \end{tabular}
 \end{footnotesize}
  \caption{Qualitative analysis of the literature reviews related with digital forensic guidelines, frameworks, tools, and readiness. Notation: \checkmark denotes that this topic is analysed, while $\circ$ denotes that its only partially discussed or just named.}
 \label{tab:frameworks_global_challenges}
\end{table*}%

\subsection{Forensic readability and reporting}
\label{sec:forensic_readability}


The continuous appearance of novel ICT technologies, paired with discovering new vulnerabilities and attacks that threaten them, dramatically increases the amount of information collected during forensic investigations. The latter refers to the amount of data collected from devices and systems, as well as the heterogeneous data structures required in each case and the specific forensic methodologies developed to detect such threats. In this context, creating interoperable and auditable forensic procedures is a hard task, especially due to the lack of standardised reporting mechanisms. Moreover, qualitative aspects such as the outcomes and conclusions supported by the forensic analysis are often not reported accurately in an attempt to balance between technicality and comprehensibility, hindering the robustness of the findings \cite{bali2020communicating,howes2017discord,overill2021quantitative}.
Of particular relevance is the communication and readability of such reports, especially if these are to be interpreted by law practitioners, judges, and other stakeholders who do not always have the necessary technical background about the forensic tools nor the underlying technologies analysed \cite{HOWES201454,howes2013forensic}. The latter issue has been extensively analysed according to different approaches, from lexical density and complexity \cite{halliday1989some,eggins2004introduction,flesch1948new,flesch1949art,kincaid1975derivation,clerehan2005linguistic,mosenthal1998new}, to cognitive and psychological features \cite{calderon2008applying,graves2003assessing}, showcasing the need to improve the reporting mechanisms and the possible benefits of a common, standardised framework. In addition to such a framework, it is crucial to develop the corresponding training procedures for its adoption \cite{cosic2017formal}.

It is necessary to recall that the admissibility of a piece of evidence and the forensic validation in court is mandatory to the proper prosecution of perpetrators and constitute the culminating point of an investigation \cite{sallavaci_13,sommer_11}. Therefore, several authors collected the challenges and issues related to the acceptance of evidence in court \cite{arshad_18,sallavaci_13,butler_16}. Moreover, region-focused studies can be found in \cite{sommer_11} and \cite{butler_16} for the United Kingdom and Australia, respectively.

After analysing the previous literature of forensic reporting procedures and studying the technical level of the data to be included \cite{legalexecutiveinstitute16,bariki2010defining}, as well as analysing existing investigation models such as ISO/IEC 27043:2015 \cite{karie2019importance}, we identified a set of key points and structural features that such document should include. In parallel, we analysed the technical level associated with each characteristic as reported in the literature and created a reporting guideline document, which is represented in Table~\ref{tab:reporting}. As it can be observed, summaries, overview descriptions and listings should be performed in a comprehensive, non-technical way. In the case of tool descriptions, as well as proofs guaranteeing the outcomes, the report should contain some technical yet understandable descriptions. Finally, the scientific aspects and details behind the analysis and the corresponding methodologies require descriptions that should be provided by qualified experts.

\begin{table}[t]
 \centering
 \begin{footnotesize}
 \rowcolors{2}{gray!25}{white}
 \begin{tabular}{cp{.65\linewidth}P{.125\linewidth}}
\toprule \textbf{Step} & \textbf{Description} & \textbf{Technical level}\\
 \midrule
 1 & Summary of contents & Low \\
 2 & Case information, description and examiners & Low \\
 3 & Forensic tools, versions, and main purpose of each tool. Limitations of each tool and scope. & Medium \\
 4 & Repository/evidence list and overview of the analysis and investigators behind such analysis. & Low \\
 5 & Summary of acquisition, seizure and analysis of evidence, and chain of custody preservation& Low \\
 6 & Technical aspects and methodology of the forensic analysis & High \\
 7 & Proof of replicability (repeated experiments led to same conclusions and are supported by data) & Medium \\
 8 & Link with other investigations, procedures and other remarks. & Medium \\
 \bottomrule
 \end{tabular}
 \end{footnotesize}
  \caption{Proposed representation of the content of a forensic report according to the inputs collected from the literature.}
 \label{tab:reporting}
\end{table}%

\subsection{Data management and Ethics}

When discussing digital forensics and respective technology readiness, the applicable regulatory frameworks should be considered as well. As seen in \cite{park_18}, integrating digital forensic readiness as a component in data protection legislation could improve actual practices across different sectors and countries.

In particular, this section highlights the regulatory requirements of working with data in Europe and in the European Union. To facilitate digital forensic readiness, tools should be developed and used in line with legal requirements, with special attention to the individual's privacy.

\myparagraph{Privacy in Europe}
States have numerous responsibilities concerning the protection of their citizens. Although the protection of privacy (in its various forms) is important, it represents but one of the duties states should fulfil \cite{klitou2014privacy}. Other prominent duties relate to the need to protect the life and property of citizens, to prevent disorder, to ensure that justice occurs where individuals have been the victim of criminal activity and to protect national security both offline and online \cite{daniels2001justice}. In modern western societies, it is often impossible to guarantee the exercise and protect such rights and in an absolute manner to all individuals all of the time due to competing interests of stakeholder groups. Respectively, privacy is only one of such values next to, e.g., security and the need for public order. To ensure security, the state likely has to take measures that may infringe upon the privacy of individuals \cite{neocleous2007security}. This entails the acquisition of data or the conduct of surveillance to prevent \textit{inter alia} acts of terrorism or crime. These activities clearly interfere with and limit the privacy of citizens but do so for desirable reasons. However, interference with such competing interests should be balanced, and the rights and freedoms of all groups in society should be respected to the greatest extent \cite{klitou2014privacy}. Respectively, the need to balance the privacy and security interests implies that security measures that infringe upon individual privacy are not acceptable unless they really are intended to meet a need that is relating to the protection the rights and interests of others. Where such justification does not exist, infringement of individual privacy would not be acceptable.

\myparagraph{Data protection in Europe}
In consonance with the individual's data protection interest and society's own protective endeavours toward fighting crime and securing national security, the Council of Europe and European Union developed a common framework to be observed by technology developers, security agencies, including Police, and criminal justice system.
The most relevant instruments of the Council of Europe relating to the processing of data as evidence are:
\begin{enumerate*}
 \item the European Convention for the Protection of Human Rights and Fundamental Freedoms (ECHR) in particular with reference to the protection of the rights to privacy and due process,
 \item the Council of Europe Convention on Cybercrime, as this Convention remains the main and only international treaty which defines the substantive elements of cybercrimes \cite{treaty185},
 \item the Council of Europe Convention on Mutual Assistance in Criminal Matters, and its 1978 Protocol \cite{treaty030}, and
 \item the Electronic Evidence Guide \cite{16803028af}.
 \end{enumerate*}

A second protocol concerning the ``Enhanced international cooperation on cybercrime and electronic evidence'' is also in development \cite{1680795713}.

In European Union Art. 4 (2) of the Treaty on the European Union (TEU) states that national security is the sole responsibility of each Member State. To facilitate a harmonized approach to national security, the EU adopted several Directives and other legislative pieces in connection with criminal matters such as :
\begin{enumerate*}
 \item Charter of Fundamental Rights of the European Union, art 7 and 8.
 \item 2016/679 General Data Protection Regulation
 \item Statement of the Article 29 Working Party, Data protection and privacy aspects of cross-border access to electronic evidence, Brussels, 29 November 2017.
 \item 2016/680/EU Law Enforcement Directive \cite{3A32016L0680}
 \item 2014/41/EU European Investigation Order Directive
 \item EU 2000 Convention on mutual assistance in criminal matters
 \item 910/2014 eIDAS Regulation \cite{2014.257.01.0073}
 \item Electronic evidence - a basic guide for First Responders Good practice material for CERT first responders by ENISA, and
 \item E-evidence package \cite{eevidence2019}
\end{enumerate*}

To rationalize the functioning and limit the increasing number of legal provisions, Regulation 2016/95 repealed certain acts in the field of police cooperation and judicial cooperation in criminal matters \cite{3A32016R0095}.
LEAs performing digital forensics have confidentiality case levels depending on the severity of the crime. The forensic examiners sign a special confidentiality agreement regarding data protection upon their employment. There are policies regarding data protection, all the case relevant data is kept only to the internal network, which is protected with the use of all the necessary measures (Secure Connections, encryption, controlled access at the physical location). The forensic examination equipment is not connected to the internet when examinations are conducted.
The data in question in digital forensics is referred to as electronic evidence, defined as ``any information (comprising the output of analogue devices or data in digital format) of potential probative value that is manipulated, generated through, stored on or communicated by any electronic device'' \cite{608185}. Respectively, to use such data, specific rules concerning the gathering and use of (digital) evidence should be adhered to as well. Electronic evidence is admissible in courts when the following sets of rules are adhered to:
\begin{enumerate*}
 \item general rules and principles concerning due process in criminal proceedings;
 \item general rules of evidence in criminal proceedings and;
 \item specific rules relating to electronic evidence in criminal proceedings \cite{insa2007admissibility}.
\end{enumerate*}

There are both current, and to-be adopted elements of the applicable legal framework, but it must be underlined that as of now, there is no comprehensive international or European legal framework providing rules relating to evidence \cite{biasiotti2018handling}.
From these documents, five overarching principles can be deducted concerning the acquisition and use of electronic evidence. These are: data integrity, audit trail, specialist support, appropriate training, and legality \cite{anderson2015electronic}. National criminal procedure codes (referred above) contain further, specific provisions regarding the record and applicability of digital evidence in criminal procedures.

\section{Discussion}
\label{sec:discussion}


In Section \ref{sec:taxonomy}, we provided a topic-based taxonomy of the digital forensics literature. In what follows, we recall the challenges identified in each category and provide some strategies to overcome them.

\subsection{The road ahead in digital forensics' topics}

After revising the challenges collected in cloud forensics, most of them are closely related to data management. More concretely, data acquisition, logging, limited access to forensic data, cross-border data access and exchange are vital parameters in cloud forensics. In terms of log management, Marty et al. \cite{Marty2011} proposed using log management architecture and the guidelines for application logging in SaaS service model using technologies such as Django, Javascript, Apache, and MySQL. A centralised logging scheme was proposed by Trenwith et al. \cite{Trenwith2013} to accelerate the investigation process and provide forensic readiness. Patrascu et al. \cite{patrascu2014} proposed a scheme to monitor various parallel activities in a cloud environment. In addition to the previous works, several authors have devoted efforts towards efficient and secure evidence management in the cloud \cite{kebande2015,zawoad2015,ahsan2018}, including the use of blockchain such as seen in \cite{wang2019}. We believe that efficient evidence and logging collection mechanisms paired with secure and verifiable management of such evidence are crucial to guarantee sound cloud forensic investigations.

Network traffic forensics is a long-standing domain with numerous research efforts and tools. The main gaps that currently exist and on which future efforts shall be focused are related to the volume of the traffic, the different protocols that emerge mainly due to the IoT rise, and the fact that traffic is encrypted in most cases. As the use of computer systems and the internet grows exponentially, the network traffic size to be analysed to conduct a forensics investigation rises. Methods that can efficiently analyse voluminous traces of network traffic are in high demand. Additionally, the heterogeneity of network traffic protocols increases the effort required to collect evidence from all available sources.

Last but not least, the main challenge that network forensics research faces nowadays is encrypted traffic. When digital forensic evidence acquisition happens at an intermediate node of the communication path, it is expected for the traffic payload to be encrypted, and methods capable of extracting information under such conditions are required.

Filesystems, Memory, and Data Storage forensics have attracted the research community's attention, as they are an abundant source of digital evidence. As discussed in Section \ref{sec:filesystems}, the main challenge of these domains lies in the fact that there exist a large number of files and data contained in them. Thus, the efforts should focus on big data analysis and data mining techniques to extract the relevant investigation data from the vast amount of unrelated or redundant digital objects. Another issue is the case of distributed filesystem and databases or data stores, or when the forensic analysis should be conducted on the cloud. In the latter case, besides the specialised tools and methods, it also challenges collaboration and cooperation with the cloud service providers. Finally, most research works and tools are bound to specific system architecture, OS, or hardware implementation, so they have the drawback of becoming cumbersome to adjust existing solutions to new use cases and problems. In this context, more generic approaches that allow tool reuse in different cases are necessary.

The recovery of digital evidence from portable and/or mobile devices is the focus of mobile forensics (MF), a sub-branch of digital forensics. Seizure, acquisition, and examination/analysis are the three categories that mobile forensics processes fall into. Several challenges exist concerning mobile forensics, as presented in \ref{sec:Mobile}. In the MF domain, the variety of embedded OSs with shorter product life cycles and the numerous smartphone manufacturers worldwide present significant challenges for applying sound forensics approaches. MF, in general, present a variety of challenges such as problems with data (anonymity-enforced browsing and other anonymity services, and the considerable volume of data acquired during an investigation), availability of forensic tools (MF research approaches have long focused on acquisition techniques, while minor importance was given to the other phases of MF investigative process) and security-oriented concerns (development of new and more sophisticated anti-forensic methods from mobile manufacturers). It is worth noting that MF is confronted with significant challenges regarding the overall MF processes' focus. For example, it is unclear whether investigation procedures should be model-specific for each device or generic enough to form a standardized set of forensics procedures guidelines. Another critical issue is the requirement to perform live forensics (mobile devices should be powered on). Finally, due to the security features built into modern mobile devices, an investigator must break into the device using an exploit that will almost certainly alter the data.

While the widespread adoption of IoT devices and IoT-related applications has improved data availability and operational excellence, it has also introduced new security and forensics challenges. As presented in Section \ref{sec:IoT}, several challenges exist concerning IoT forensics. Such challenges include managing multiple streams of data sources, the complicated three-tier architecture of IoT and the lack of standardized systems for capturing real-time logs and storing them in a valid uniform form. The preparation of highly detailed reports of all information gathered, its corresponding representation, and the lack of standardized systems for capturing real-time logs also serve as barriers to establishing sound IoT-related forensics mechanisms. Data encryption trends are also posing new challenges for IoT forensic investigators, and cryptographically protected storage systems are arguably one of the most significant roadblocks to effective digital forensic analysis. Interoperability and availability issues related to the vast number of connected IoT devices, the Big Data nature of IoT forensics evidence, and IoT forensics evidence's various security storage challenges also represent significant IoT-related forensics challenges. Finally, the IoT forensics domain faces several regulatory challenges, particularly those relating to data ownership in the cloud as defined by regional laws.

As seen in Section \ref{sec:multimedia}, multimedia forensics is one of the most explored topics, according to the number of publications. Overall, while most authors focus on image forgery detection, anti-forensics is one of the most challenging problems. In this regard, more efforts should be devoted to counter anti-forensic mechanisms (i.e., as part of a global digital forensics concern) and methodologies to capture novel criminal trends with the help of sophisticated real-time object detection and classification systems. In addition, multi-layer systems and ontologies should be designed to cope with multiple threats at once, paired with the appropriate benchmarks to evaluate them. In parallel, the issues related to the vast amount of data to be processed should be minimised by proposing more efficient data storage and indexing mechanisms and introducing algorithms that can process, e.g., compressed data. Following such research paths and combining them with the proper legislation and standardisation mechanisms will improve the success of multimedia digital forensics investigations.

Blockchain forensics is a relatively new domain since blockchain technology accounts for a decade. In general, it has to be understood that the need for blockchain forensics methods is expected to grow in the coming years. As discussed in Section \ref{sec:blockchain} current efforts focus on the examination of available data on public blockchain systems. One of the main challenges encountered is to provide efficient methods to conduct such analysis. The data on public ledgers continuously grows, while the storage structure differs amongst different implementations. Developing methods and tools that can efficiently analyse data across commonly used blockchain platforms is required. Moreover, forensic analysis methods for blockchain systems' nodes will enable more thorough investigations with more detailed results for public and private blockchain systems. Finally, given the rising popularity of privacy enabled blockchain systems such as Monero or ZCash, additional effort will be required to support forensic investigations on cases that include interactions on such systems.


\subsection{Open Issues and Future Trends}
\myparagraph{Forensic readiness and reporting}
Given the continuous evolution of cybercrime and its harmful capacities, preventive strategies are paramount to fight criminal activities. The latter implies the need to reinforce digital forensic strategies at different levels, including guidelines, regulations, research and training to implement forensic readiness holistically. According to our literature analysis, one of the key points to reinforce the actual state of practice is the definition of interoperable and easy-to-adopt legislations since current ones cannot cope with the increasing sophistication and the ubiquitous nature of cybercrime. Therefore, it is crucial to devote efforts towards, e.g., interoperable cross-border models with their corresponding dissemination and training procedures, which all practitioners may adopt to accelerate investigations.
It is also relevant to stress the necessity of appropriate forensic readability and reporting. First, effective communication between all the actors involved in a forensic investigation is essential to maximise the guarantees in court. Second, the proper documentation of investigations provides valuable feedback for future investigations, enhancing forensic readiness strategies. Third, the definition of a common reporting framework can accelerate investigations in which sometimes speed is crucial due to, e.g., the possible volatility of evidence or to reduce harm. To this end, we proposed a forensic reporting content representation by following the common denominators found in the literature in Section \ref{sec:taxonomy}. We argue that the devotion of more efforts on this final part of the forensic flow will enrich investigations with valuable feedback and successful prosecution guarantees.

\myparagraph{Forensic preparedness and standards}
While in Section \ref{sec:digitalforensics_mps} we provided an overview of digital forensics standards, unfortunately, they do not suffice current needs. To name just two which are standing out on the tip of the iceberg, cloud and mobile related investigations need to have some standards on how to be performed. Addressing the need for mobile forensics, FORMOBILE\footnote{\url{https://formobile-project.eu/}} has initiated a broad dialogue and is developing a draft CENELEC Workshop Agreement to fill in this gap. However, due to the specificities of cloud, IoT, drones, etc., similar actions are expected in the near future.

Beyond standards and methods, there is a definite need from industry players, developers, system administrators etc., to foster a culture of \textit{forensic preparedness}. Essentially, every organisation and resource provider must understand that its products and services are expected to suffer a successful cyber attack. Therefore, despite the countermeasures, recovery methods, and mitigation strategies, they need to implement policies and mechanisms to facilitate digital forensics. If the latter are not well-placed, while business continuity may not be severely harmed, one may not understand why and how the security event occurred, what needs to change, or miss even important evidence of the threat actor.

\myparagraph{Decentralisation and immutability}
The wide adoption of distributed platforms, e.g. blockchain solutions \cite{casino2018systematic} and distributed storage and filesystems, imply significant challenges for digital forensics \cite{casino2019immutability,kebande2022review}. Some of these structures have strong privacy guarantees and can be leveraged to exfiltrate data, orchestrate malicious campaigns \cite{ali2018zombiecoin,ipfsmalware,bazaar,casino2021unearthing}, or siphon fraudulent payments \cite{BalthasarH17}. Traditional logging mechanisms and access control systems allow an investigator to assess who, when, how or even from where are not relevant for many of these technologies. As a result, they are continuously abused by threat actors. These huge obstacles for digital forensics require further research on the field and the development of more targeted tools to extend the capabilities of digital investigators. In this regard, while the use of distributed platforms is not exempt from potential issues \cite{kebande2022review}, they can also be potentially used to leverage community-based intelligence against threats and to leverage auditable forensic investigations\cite{dasaklis_20,kumar2021internet,locardproject,zarpala2021blockchain}. Following such an idea and in order to accelerate the response towards sophisticated threats and international campaigns, the community is devoting research efforts towards federated learning models \cite{li2020federated,yang2019federated}, and other emerging topics such as cognitive security \cite{ogiela2020cognitive,demertzis2018next}.

\myparagraph{Data protection and ethics in criminal investigations} Ransomware may be regarded as the most obvious case of exploiting cryptographic primitives for malicious acts; nevertheless, this is not by any chance the only. Threat actors and cybercriminals, for instance, use encrypted and even covert channels to communicate, further hindering investigations. The latter has sparked a huge debates as many are promoting concepts such as \textit{responsible encryption}\footnote{\url{https://www.justice.gov/opa/speech/deputy-attorney-general-rod-j-rosenstein-delivers-remarks-encryption-united-states-naval}} with the adoption of, e.g., weakened encryption, cryptographic schemes such as key escrow, backdooring of cryptographic primitives etc. \cite{SCHUSTER201776,bernstein2016dual,smith2016discussion,rice2017second}. While they may facilitate digital investigations, essentially, they undermine the scope of cryptography and security, opening the door for many interpretations on what lawful interception is, when it can be performed, by whom, let alone the exploitation of the mechanisms by already malicious actors as the backdoor would be already implanted. The debate is undergoing and spans multiple sectors beyond digital forensics. While fostering such approaches may greatly benefit digital forensics, the ethical and legal implications hinder such adoption and are received by the security community with scepticism.
As discussed, anti-forensics methods are a challenge for almost all domains of digital forensics. Nevertheless, with the growing adoption of TPM and TEE, these challenges can be significantly augmented. For instance, as illustrated by Dunn et al. \cite{266514} ransomware can exploit these technologies to render decryption key extraction impossible. Nevertheless, it is clear that these technologies introduce significant challenges for digital investigators since they may deprive them of access to critical information. In this regard, it is essential to study methods for, e.g. live forensics in the presence of TPM and TEE and to explore how the missing information can be compensated.

\myparagraph{Automation and explainability} The continuous increase in reported cybercrimes apart from the impact on the victims implies a lot of effort from investigators to analyse the cases. Therefore, automation of digital forensics inevitably becomes a need. While automated methods for collecting log files and algorithms to identify anomalies or even correlating some events may exist, this does not practically translate to automated digital forensics. Even if one does not consider APT attacks, one must understand that each case has particularities differentiating it from the others. Moreover, a digital investigator has to fill in the gaps of missing information that the attacker managed to cover, including those that security mechanisms failed to record or those erroneously reported. The above implies the development of advanced machine learning and AI algorithms and tools that will underpin future digital forensics investigations. An important part of these systems is undoubtedly understanding the scope of the investigation and the explainability of the results \cite{adadi2018peeking}, which is critical to assess the impact of current investigations and quantify their effectiveness \cite{overill2021quantitative}, a critical step to ensure the implementation of the proper measures. The latter is a crucial part of AI and machine learning modules that have to be introduced as in order for a piece of evidence to be admissible in a court of law, one has to justify not only how and from where it has been collected but to also prove the relevance to the case, how it was used, and why it is linked with the rest of the evidence. In essence, future digital forensics systems would have to argue and reason on the collected information in a human-readable manner. The latter is a huge step forward compared to the existing state where systems prioritise log events and present the analysts with known malicious patterns in the logs, malicious binaries, or connections that deviate specific norms.


\myparagraph{Forensic guidelines and best practices} One of the main strategies to reduce the impact of cybercrime is to implement the recommendations of the security guidelines and directives developed by agencies such as ENISA and NIST. The current threat landscape \cite{enisa2021threat}, which includes ransomware, malware, and threats against data availability and veracity, affect digital forensics in different dimensions, regardless of the topic. NIST recently published a state of the art analysis of cloud-related challenges \cite{nist2020cloud}, which is aligned with the claims collected by in the cloud-based digital forensics literature reviews state in Section \ref{sec:cloud_taxonomy}.
In the case of networks, ENISA elaborated an extensive set of security objectives and discussed them along with their corresponding recommendation measures in the topics of electronic communications \cite{enisa2020eecc} as well as 5G networks \cite{enisa20205g}.
NIST provides security guidelines for managing mobile devices in their draft SP 800-124 (rev2) \cite{nist2020mobile}. The recommendations include scenarios from organization-provided to personally-owned devices and describes technologies and strategies that can be used as countermeasures and mitigations. In the context of IoT, NIST released a set of documents related to IoT device cybersecurity, covering aspects from the design and manufacturing of the components to their disposal \cite{nist2020iot}. In parallel, ENISA also proposed a comprehensive set of security guidelines targeting all the entities involved in the supply chain of IoT to improve security decisions when designing, building, deploying, and assessing IoT technologies \cite{enisa2020iot}.
Concerning data storage and data processing, several guidelines have been proposed during the past years to reduce data breaches \cite{enisa2020pdp}, and the proper deployment of data storage mechanisms that enable privacy by design \cite{nist2020datapriv,enisa2020pbdf,zigomitros2020survey}, and forensic readiness \cite{enisa2020auage}. Finally, despite the existence of such guidelines, the adoption of methodologies that enable the review and evaluation of an investigation process is critical to assessing the quality of investigations and their improvement \cite{sunde2021part}.

\section{Conclusions and final remarks}
\label{sec:conclusions}

The digitisation of our daily lives is a double-edged sword as beyond the myriad of advantages and comforts it provides, it introduces security and privacy issues. Motivated by the lack of a general view of the digital forensics ecosystem, mainly because different topics are explored in an isolated way and aiming to answer several research questions/concerns, this manuscript seeks to fill a literature gap by proposing a review of reviews in the field of digital forensics. Following a thorough research methodology, we identified the main digital forensics topics. We performed a taxonomy by documenting the current state of the art and practice and the main challenges in each of them. Moreover, we analysed these challenges with a cross-domain perspective to highlight their relevance according to the times they were discussed in the literature. Such analysis provided us with enough evidence to prove that the digital forensics community could benefit from closer collaborations and cross-topic research since it appears that researchers are trying to find solutions to the same problems in parallel, sometimes without noticing it.

By merging the information of Table \ref{tab:cross_limitations} and Figure \ref{fig:circles}, we extracted the amount of cross-domain challenges that each topic has in each forensic phase, and reported them in Table \ref{tab:lim_per_level}. As it can be observed, data acquisition along with investigation and forensic analysis are the phases that entail more challenges, according to the research community. If we analyse the data at a topic level, we can observe that IoT has many challenges to overcome in such phases. The same applies to Multimedia and Mobile forensics. Since we focus on the extracted challenges as collected in our literature review, the fact that some challenges have not been highlighted either at topic or forensic phase level may indicate that researchers and practitioners have not devoted enough effort to them, or perhaps highlights lack of discussion towards them. Such interesting domains include value chain and financial forensics. Like other domains, the business sector's ongoing digitisation means that sound value chain forensics mechanisms will be almost necessary within any corporate strategy for the years to come. Therefore, the potentially unexplored issues in such cases require proactive initiatives before they become obstacles in the near future.

\begin{table}[hbt!]
\renewcommand{\tabcolsep}{1mm}
 \centering
 \scriptsize
 \rowcolors{2}{gray!25}{white}
\begin{tabular}{lP{1.5cm}P{1.85cm}P{1.75cm}P{1.5cm}}
\toprule
 & \textbf{Standards \& legislation } & \textbf{Data acquisition \& pre-processing} & \textbf{Investigation \& forensic analysis } & \textbf{Reporting \& presentation}  \\
 \midrule
\textbf{Cloud }   & 2     & 3      &   2 &        1   \\
\textbf{Networks }&       &  2     &     1   &        \\
\textbf{Mobile}   &2      & 3     &   3 &             \\
\textbf{IoT }     & 2     & 4      &  5  &          1  \\
\textbf{FS \& DB} &1      & 3      &    2   &      1    \\
\textbf{Blockchain} &     &   2   &      1    &        \\
\textbf{Multimedia}  &    &  4     &   3    &         \\

 \bottomrule
\end{tabular}
\caption{Limitations per topic according to each phase as depicted in Figure \ref{fig:circles}}
 \label{tab:lim_per_level}%
\end{table}

Further to merely listing the state of practice and proposing research directions according to the identified challenges, we analysed crucial aspects of digital forensics such as standards, forensic readiness, forensic reporting, and discussed the ethical and legal aspects of data management in Europe. The insights gathered from such analysis were represented in the form of structured tables, qualitative literature analysis, and a proposed representation of forensic report content.

Finally, we discussed the main takeaways of this article and showcased several challenges that the digital forensics community will face in the upcoming years. In this regard, we proposed some ideas to prevent and/or overcome them while recalling the need to design efficient and cross-domain strategies since the latter will guarantee faster and more robust outcomes, hopefully minimising the impact of criminal activities.

The inherent cross-jurisdiction nature of modern cybercrime paired with the abuse of cutting edge technologies mandates more coordinated efforts from the security and research community. With the continuously increasing amount of data that have to be analysed, it is straightforward that manual analysis is almost at its limits. The use of fine-grained IoCs may significantly reduce the effort of the investigator. However, as already discussed, this is not always possible, especially when non-traditional computing devices are used, e.g. IoT, mobile, cloud. As a result, machine learning and artificial intelligence are gradually being integrated into the logic of many tools and methods. Nevertheless, the reasoning of the results in an understandable human manner is a cross-domain challenge. Moreover, the standardisation of digital forensics processes for cloud, mobile, IoT, drones, etc., is becoming a high priority since they are an indispensable part of almost all modern digital investigations. Finally, the consensus on developing these standards and the coordinated efforts made over the past few years for countering cybercrime must be leveraged to homogenise the legislation across jurisdictions and facilitate digital investigations. A common answer to the problem and using the same measures would create a strong response against cybercrime and improve response time to security incidents and their analysis.

\section*{Acknowledgement}
This work was supported by the European Commission under the Horizon 2020 Programme (H2020), as part of the projects  \textit{LOCARD} (\url{https://locard.eu}) (Grant Agreement no. 832735) and CyberSec4Europe (\url{https://www.cybersec4europe.eu}) (Grant Agreement no. 830929).

F. Casino was supported by the Beatriu de Pinós programme of the Government of Catalonia (Grant No. 2020 BP 00035).

\end{document}